\documentclass{pasj00}

\begin{document}
\SetRunningHead{K. Morokuma-Matsui et al.}{Stacking Analysis of $^{12}$CO and $^{13}$CO Spectra of NGC~3627}

\title{Stacking Analysis of $^{12}$CO and $^{13}$CO Spectra of NGC~3627:\\ Existence of non-optically thick $^{12}$CO emission?}

\author{Kana \textsc{Morokuma-Matsui}} %
\affil{Nobeyama Radio Observatory, 462-2 Nobeyama, Minamimaki-mura, Minamisaku-gun, Nagano 384-1305}
\email{kana.matsui@nao.ac.jp}

\author{Kazuo \textsc{Sorai}}
\affil{Department of Physics/Department of Cosmosciences, Hokkaido University, Sapporo, Hokkaido 060-0810}
\email{sorai@astro1.sci.hokudai.ac.jp}

\author{Yoshimasa \textsc{Watanabe}}
\affil{Department of Physics, The University of Tokyo, 7-3-1, Hongo, Bunkyo-ku, Tokyo 113-0033}
\email{nabe@taurus.phys.s.u-tokyo.ac.jp}

\and
\author{Nario {\sc Kuno}}
\affil{Faculty of Pure and Applied Sciences, University of Tsukuba, 1-1-1, Tennoudai, Tsukuba, Ibaraki 350-8577}
\affil{Nobeyama Radio Observatory, 462-2 Nobeyama, Minamimaki-mura, Minamisaku-gun, Nagano 384-1305}
\affil{The Graduate University for Advanced Studies (SOKENDAI), 2-21-1 Osawa, Mitaka, Tokyo 181-8588}
\email{kuno.nario.gt@u.tsukuba.ac.jp}


%


\maketitle

\begin{abstract}
We stacked $^{12}$CO and $^{13}$CO spectra of NGC~3627 after redefining the velocity axis of each spectrum of the mapping data so that the zero corresponds to the local mean velocity of $^{12}$CO spectra. 
The signal-to-noise ratios of the resulting spectra are improved by a factor of up to 3.2 compared to those obtained with normal stacking analysis.
We successfully detect a weak $^{13}$CO emission from the interarm region where the emission was not detected in the individual pointings.
We compare the integrated intensity ratios $I_{^{12}\rm{CO}}/I_{^{13}\rm{CO}}$ among six characteristic regions (center, bar, bar-end, offset, arm, and interarm).
We find that $I_{^{12}\rm{CO}}/I_{^{13}\rm{CO}}$ in the bar and interarm are higher than those in the other regions by a factor of $\sim2$ and $I_{^{12}\rm{CO}}/I_{^{13}\rm{CO}}$ in the center is moderately high.
These high $I_{^{12}\rm{CO}}/I_{^{13}\rm{CO}}$ ratios in the bar and center are attributed to a high intensity ratio ($T_{^{12}\rm{CO}}/T_{^{13}\rm{CO}}$) and one in the interarm is attributed to a high ratio of the full width at half maximum of spectra (FWHM$_{^{12}\rm{CO}}/$FWHM$_{^{13}\rm{CO}}$).
The difference between FWHM$_{^{12}\rm{CO}}$ and FWHM$_{^{13}\rm{CO}}$ of the interarm indicates the existence of two components, one with a narrow line width ($\sim$ FWHM$_{\rm ^{13}CO}$) and the other with a broad line width ($\sim$ FWHM$_{\rm ^{12}CO}$).
Additionally, the $T_{^{12}\rm{CO}}/T_{^{13}\rm{CO}}$ ratio in the broad-line-width component of the interarm is higher than the other regions.
The high $T_{^{12}\rm{CO}}/T_{^{13}\rm{CO}}$ in the center and bar and of the broad-line-width component in the interarm suggest the existence of non-optically thick $^{12}$CO components.
We find that more than half of the $^{12}$CO emissions of the interarm are likely to be radiated from the diffuse component.
Our result suggests that the use of a universal CO-to-H$_2$ conversion factor might lead to an overestimation of molecular gas mass and underestimation of star-formation efficiency in the interarm by a factor of a few.
\end{abstract}

\section{Introduction}

Molecular clouds are birthplaces of stars that form the visible backbones of galaxies.
Therefore, it is important to investigate the molecular gas mass, its distribution in galaxies and its physical properties to understand galaxy evolution.
Molecular gas in galaxies mainly consists of H$_2$ molecules.
However, H$_2$ molecules do not radiate line emissions in cold environments such as molecular clouds whose temperature is typically a few tens K, since H$_2$ is a homonuclear diatomic molecule (i.e., no dipole moment).
Molecular gas mass is usually estimated indirectly from observations of heteronuclear diatomic molecules such as $^{12}$CO and its isotopologues since they are the second most abundant molecules after H$_2$ in the interstellar medium.
Therefore, $^{12}$CO$(J=1-0)$ line ($^{12}$CO, hereafter) has been commonly used to observe molecular gas in external galaxies as well as the Galactic objects.

It has been observationally shown that the $^{12}$CO line luminosity correlates with the molecular gas mass estimated in several different ways, such as virial techniques, dust emission, extinction mapping and gamma-ray observation (\cite{Bolatto+2013} and references therein).
The correlation between $^{12}$CO luminosity and molecular gas mass is explained theoretically under the assumptions that 1) molecular clouds are virialized, 2) the mass of clouds are dominated by H$_{2}$, 3) the clouds follow the size-line width relation and 4) they have a constant temperature (\cite{Bolatto+2013} and references therein).
According to the observational evidence and the theoretical explanation above, the mass and distribution of molecular gas in nearby galaxies have been investigated with the observed $^{12}$CO maps by assuming a constant CO-to-H$_2$ conversion factor over a whole galaxy (e.g., BIMA SONG, \cite{Helfer+2003}; Nobeyama CO atlas, \cite{Kuno+2007}, hereafter K07; Heracles, \cite{Leroy+2009}).

For a more accurate estimation of the molecular gas mass, multiple CO-isotopologues should be used in a complementary manner.
The mass of virialized molecular clouds can be measured with $^{12}$CO and the higher opacity of $^{12}$CO allows us to trace the molecular gas with lower densities on the order of $10^2$ cm$^{-3}$.
On the other hand, the opacity of $^{12}$CO is too high to estimate the column density through the molecular cloud unlike more optically thin lines such as $^{13}$CO$(J=1-0)$ and C$^{18}$O$(J=1-0)$ ($^{13}$CO and C$^{18}$O, hereafter).
However, the line intensities of these CO-isotopologues are usually much weaker than that of $^{12}$CO (by factors of $\sim1/5-1/20$ for $^{13}$CO and \hspace{0.3em}\raisebox{0.4ex}{$<$}\hspace{-0.75em}\raisebox{-.7ex}{$\sim$}\hspace{0.3em}$1/20$ for C$^{18}$O, respectively) and require a lot of telescope time to be detected.
Therefore, $^{13}$CO or C$^{18}$O mapping observations toward nearby galaxies have been limited to only a handful of cases \citep{Huttemeister+2000,Paglione+2001,Tosaki+2002,Hirota+2010,Watanabe+2011}.

Some studies have investigated the relation between the physical states of molecular gas and galactic structures such as arms and bars from the CO multi-isotopologue observations (e.g. \cite{Huttemeister+2000,MeierTurner2004}).
\citet{Watanabe+2011} (hereafter W11) observed NGC~3627 in $^{13}$CO and showed high $^{12}$CO/$^{13}$CO intensity ratios in the bar region.
They concluded that the high $^{12}$CO/$^{13}$CO ratio indicates the existence of gravitationally unbound diffuse gas as a result of the strong streaming motion in the bar region.
Spiral arms are also likely to affect not only the dynamics of molecular clouds but also their internal physical conditions.
The molecular gas in spiral galaxies is expected to be accumulated at the arm region through galactic shocks (e.g. \cite{Fujimoto1968,Roberts1969,Egusa+2011}) and sheared out in interarm regions due to the differential rotation.
However there are only a few studies that focus on the physical states of molecular gas in interarm regions since even the $^{12}$CO emission is quite weak \citep{Tosaki+2002}.
To understand the effects of spiral arms on the molecular clouds, it is necessary to detect emissions of CO isotopologues in the first place and then investigate the physical states of molecular gas in the interarm region.

The aim of this paper is to detect weak $^{13}$CO emission and investigate the relationship between the galactic structures and properties of molecular gas by comparing $^{12}$CO and $^{13}$CO spectra in the characteristic regions of spiral galaxies.
We stacked $^{12}$CO and $^{13}$CO spectra of a nearby barred spiral galaxy, NGC~3627\footnote{
NGC~3627 is classified as SABb in the Third Reference Catalog of Bright Galaxies (RC3, \cite{deVaucouleurs+1991}) and has slightly asymmetric spiral arms.
This asymmetric feature of the spiral arms is thought to have arisen from a past interaction with a neighboring galaxy, NGC~3628 (\cite{Haynes+1979,Zhang+1993}).
}
obtained in previous studies (K07\nocite{Kuno+2007} and W11\nocite{Watanabe+2011}) after shifting the velocity axis so that the zero of the $^{13}$CO spectra corresponds to the local mean $^{12}$CO velocity.
This stacking method was originally proposed by \citet{Schruba+2011} where the local mean velocity of HI is adopted as the zero velocity of $^{12}$CO spectra in the outer region of galaxies.
This allows us to improve the signal-to-noise ratios (S/N) of the spectra.
Then we discuss the physical properties of molecular gas in the interarm region by comparing the $^{12}$CO and $^{13}$CO spectra.

The structure of this paper is as follows:
The data and method are explained in section \ref{DandA}.
We show the results of the stacking analysis of the $^{12}$CO and $^{13}$CO spectra in the different regions of NGC~3627 in section \ref{Results}.
We compare the $^{12}$CO with $^{13}$CO stacked spectra in section \ref{Analyses} and discuss the physical properties of the molecular gas in the different regions of NGC~3627 in section \ref{Discussion}.
Finally, we summarize this study in section \ref{Summary}.

\section{Data \& stacking technique}\label{DandA}

We first summarize the data we used and the stacking method in the following sub-sections.

\subsection{Data}

\begin{table*}
\caption{Summary of the $^{12}$CO and $^{13}$CO observations.}
\begin{center}
\begin{tabular}{lcc}
\hline
&$^{12}$CO&$^{13}$CO\\
\hline
\hline
Date & April in 2004 & May in 2007 and April in 2008\\
Telescope & the 45-m telescope at NRO & the 45-m telescope at NRO\\
$\eta_{\rm mb}$ & 0.4 & 0.31\\
Receiver & BEARS & BEARS\\
Backend & digital spectrometers & digital spectrometers\\
Band width (MHz) & 512 & 512\\
Grid spacing & $10''.3$ & $10''.3$\\
r.m.s. (mK, $T_{\rm mb}$) & $40-100$ & $6-16$\\
Velocity resolution (km s$^{-1}$) & 5 & 20\\
Reference & \cite{Kuno+2007} (K07) & \cite{Watanabe+2011} (W11)\\
\hline
\end{tabular}
\end{center}
\label{tab-1}
\end{table*}%

$^{12}$CO and $^{13}$CO mapping data of NGC~3627 were both obtained with the 25-BEam Array Receiver System (BEARS) which is a 25 $(5\times5)$ beam SIS receiver mounted on the 45-m radio telescope at the Nobeyama Radio Observatory (NRO) (K07\nocite{Kuno+2007}; W11\nocite{Watanabe+2011}).
The rest-frame frequencies of $^{12}$CO and $^{13}$CO are adopted as $115.27120$ GHz and $110.20135$ GHz, respectively.
For the backend, 25 digital spectrometers \citep{Sorai+2000} were used with a total bandwidth of 512 MHz and frequency resolution of 605 kHz centered on the frequency corresponding to the local standard of rest (LSR) receding velocity.
We adopted $16''$ as a half-power beam width (HPBW) of both data in the same way as W11\nocite{Watanabe+2011}, which corresponds to $\sim$ 800 pc assuming a distance of 11.1 Mpc to NGC~3627 \citep{Saha+1999}\footnote{To be exact, the beam sizes at the frequency of $^{12}$CO and $^{13}$CO are $\sim16''$ and $\sim17''$, respectively but no correction is applied since the error due to this is expected to be small, especially for the spatially extended sources ($\sim10\%$ even for a point source, W11\nocite{Watanabe+2011}).}.
The grid spacing of the map is $10''.3$ ($\sim550$ pc).
The typical r.m.s. noise temperatures (in $T_{\rm mb}$ scale) of $^{12}$CO and $^{13}$CO data are $40-100$ mK at 5 km s$^{-1}$ resolution and 6$-$16 mK at 20 km s$^{-1}$ resolution, respectively.
The profile maps of $^{12}$CO and $^{13}$CO are shown in figure \ref{fig:ProfileMap}.
The observations are summarized in table \ref{tab-1}.
More detailed information of the observations is described in the original studies, K07\nocite{Kuno+2007} for $^{12}$CO and W11\nocite{Watanabe+2011} for $^{13}$CO, respectively.

\subsection{Stacking analysis of $^{12}$CO and $^{13}$CO spectra with Velocity-axis Alignment (VA)}

Schruba et al.(2011, 2012\nocite{Schruba+2011,Schruba+2012}) improved the sensitivity for $^{12}$CO emission in outer regions of galaxies by up to about one order of magnitude over previous studies with the stacking method they invented.
One way to reduce the r.m.s. noise temperature is to stack the spectra from various regions in the galaxy.
However, because the systemic velocities of each part of the galaxy are different due to galactic rotation, simple stacking will result in a smeared spectrum and may not yield the highest S/N.
The method adopted by \citet{Schruba+2011} overcomes this problem by shifting the spectrum in the velocity axis so that they are aligned with the mean of the local HI velocity, before stacking the spectra.
The S/N of the integrated intensity increases by a factor of $\sqrt{\Delta V_{\rm emi}/\Delta V_{\rm emi, VA}}$, taking into account that the error of the integrated intensity can be expressed as $\Delta T \sqrt{\Delta V_{\rm emi} \Delta v}$, where $\Delta V_{\rm emi}$ is the velocity range to be integrated to calculate the integrated intensity of the spectra obtained with normal stacking, $\Delta V_{\rm emi, VA}$ is one with stacking after the velocity-axis alignment, $\Delta T$ is the r.m.s. noise temperature and $\Delta v$ is the velocity resolution of the data \citep{Schruba+2012}.
Here $\Delta V_{\rm emi, VA}$ is narrower than $\Delta V_{\rm emi}$ thanks to the velocity-axis alignment.
Another reason is the reduction of frequency-dependent noise produced by systematic effects of weather, receiver instabilities and standing waves occurring during the transmission of the signal.
These sources of additional noise are canceled out if the spectra are stacked after applying different velocity shifts for each pixel \citep{Schruba+2011}.

\citet{Schruba+2011} stacked $^{12}$CO spectra of the outer HI-dominated region after shifting the velocity axis so that the zero velocity of the $^{12}$CO spectra corresponds to the local mean HI velocity.
In this paper, we adopt an intensity weighted mean velocity of $^{12}$CO as a reference velocity for the velocity-axis alignment procedure and apply the stacking method of \citet{Schruba+2011} to improve the S/N of $^{13}$CO spectra as well as that of $^{12}$CO spectra. 

\begin{figure*}
\includegraphics[width=168mm]{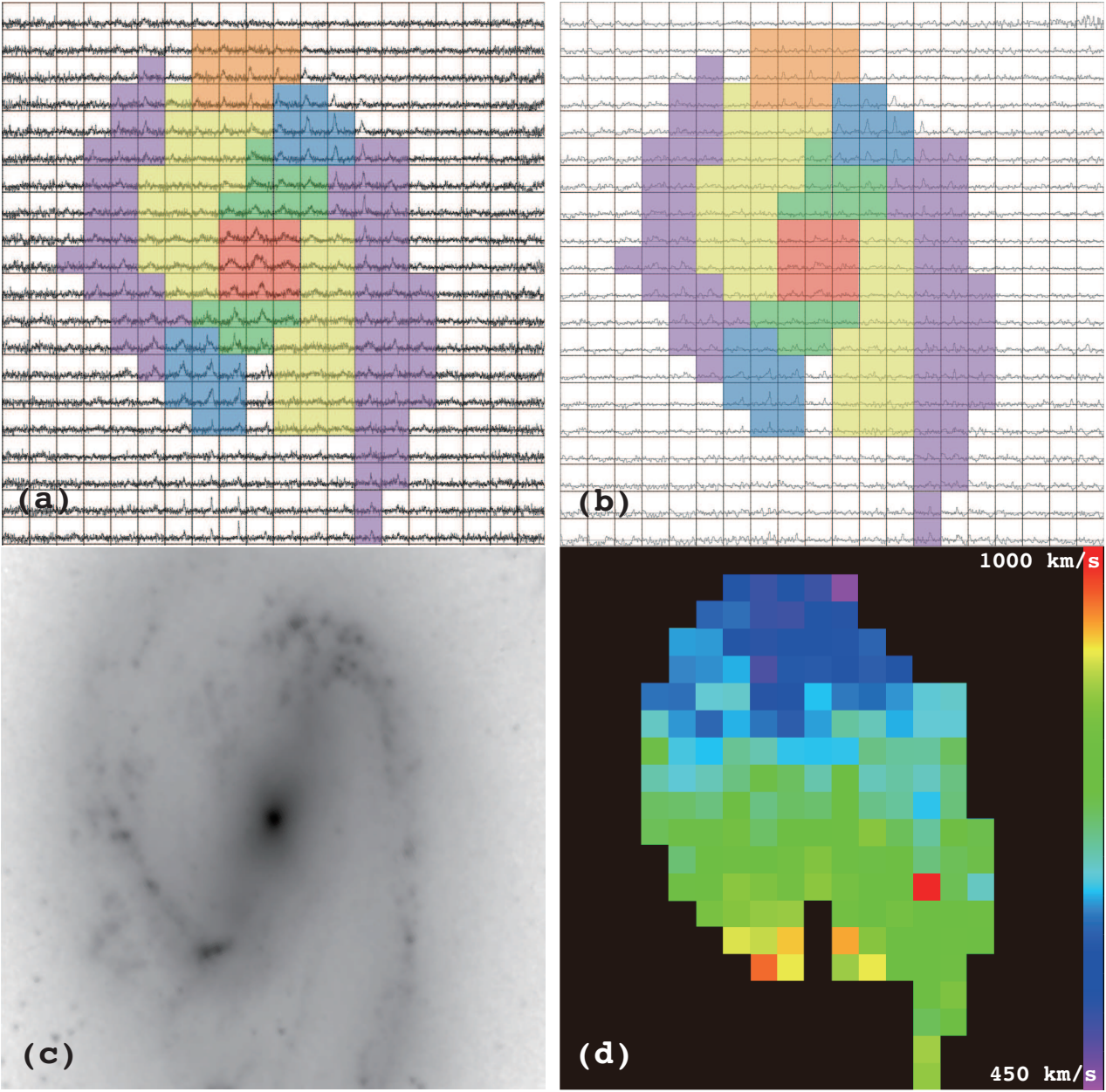}
 \vspace{0cm}
 \caption{(a) $^{12}$CO and  (b) $^{13}$CO profile maps of NGC~3627 obtained with the 45-m telescope at NRO (K07\nocite{Kuno+2007}; W11\nocite{Watanabe+2011}). The mapping area is $3'.2\times3'.2$ centered on $(\alpha, \delta)_{\rm J2000} = (11^{\rm h}20^{\rm m} 15^{\rm s}.027, +12^\circ 59' 29''.58)$. The grid size of both data are $10''.3$ and $\sim550$ pc in a linear scale. The center, bar, bar-end, offset, arm and interarm regions are colored red, green, blue, orange, purple, and yellow, respectively. (c) Spitzer/IRAC $3.6$ $\mu$m image \citep{Kennicutt+2003}. (d) 1st-moment map of the $^{12}$CO emission.}
  \label{fig:ProfileMap}
\end{figure*}

First, the velocity field of NGC~3627 is estimated with the $^{12}$CO data.
The intensity-weighted mean velocity of each pixel of the $^{12}$CO map is given by,
\begin{equation}
\overline{v_{\rm ^{12}CO}} = \frac{\int v T_{\rm mb}(v) dv}{\int T_{\rm mb}(v) dv}.
\end{equation}
The obtained first-moment map is shown in figure \ref{fig:ProfileMap} (d).
Then the shifted velocity $v_{\rm VA}$ of each spectrum is defined as
\begin{equation}
v_{\rm VA} = v_{\rm LSR} - \overline{v_{\rm ^{12}CO}},
\end{equation}
where $v_{\rm LSR}$ is the original velocity of the spectrum.
Hereafter, we refer to this procedure as velocity-axis alignment (VA).

The averaged spectra of six different regions, 1) center, 2) bar, 3) bar-end, 4) offset\footnote{Offset region denotes an area where the emission runs off toward the leading side of the stellar bar (W11\nocite{Watanabe+2011}).}, 5) arm and 6) interarm were obtained by stacking the spectra after the VA procedure\footnote{
We integrated full velocity range for the calculation of the local mean velocities of the all regions but the offset region.
The mean velocities of the spectra in the offset region, which is shown as an orange region in figure \ref{fig:ProfileMap}, are estimated by the integration ranging from $487.5$ to $687.5$ km s$^{-1}$, since we could not obtain adequate value in case of the full range integration due to poor quality of the baseline.}.
The regions of 1) $-$ 4) were determined according to W11\nocite{Watanabe+2011}.
We visually defined 5) arm and 6) interarm regions by dividing the ``other'' region in W11\nocite{Watanabe+2011} according to optical and near infrared (NIR) images.
NIR image ($3.6$ $\mu$m) from the SIRTF Nearby Galaxies Survey (SINGS, \cite{Kennicutt+2003}) is shown in figure \ref{fig:ProfileMap}c.
Each area is illustrated in different colors in figures \ref{fig:ProfileMap}a and \ref{fig:ProfileMap}b in the following way, the center in red, bar in green, bar-end in blue, offset in orange, arm in purple and interarm regions in yellow.
Finally, we averaged the velocity-axis aligned spectra with equal weights and obtained a stacked spectrum in each region.

\section{Results}\label{Results}

\begin{figure*}
\includegraphics[width=150mm]{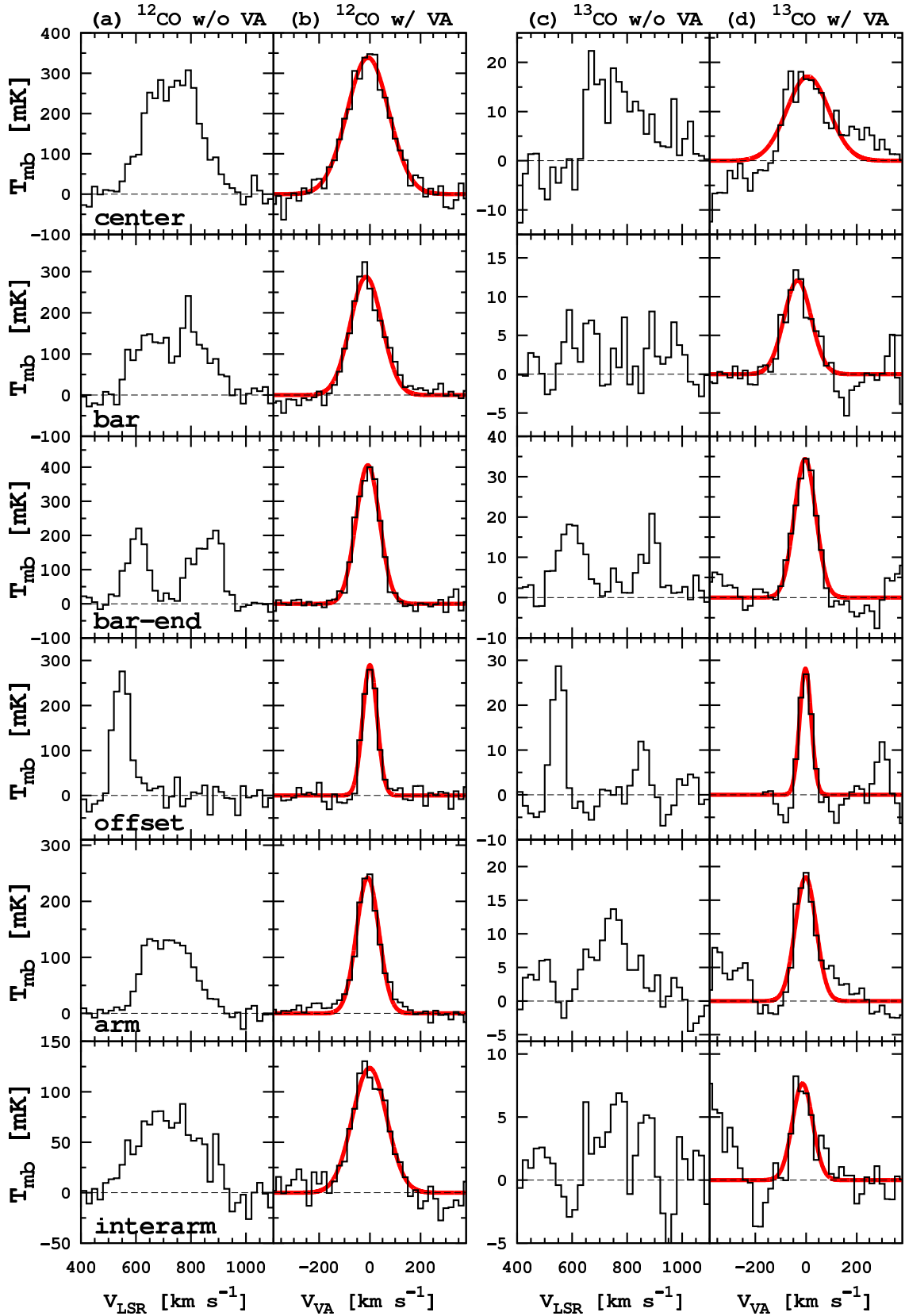}
 \vspace{0cm}
 \caption{(a) $^{12}$CO stacked spectra without the velocity-axis alignment (VA), (b) $^{12}$CO stacked spectra with VA, (c) $^{13}$CO stacked spectra without VA and (d) $^{13}$CO stacked spectra with VA.
 The vertical axis of each spectrum is main-beam temperature in mK.
 The velocity resolution of $^{12}$CO and $^{13}$CO spectra is 20 km s$^{-1}$.
 The fitting results with a Gaussian for the stacked spectra with VA are also plotted with red lines on each spectrum.
 }
  \label{fig:StackedSpectra}
\end{figure*}

\begin{table*}
 \caption{The properties of the spectrum in each different region.}
 \label{tab-2}
 \begin{minipage}{\textwidth}
 \begin{center}
 \begin{tabular}{@{}lccccc}
  \hline
   Line& $^{12}$CO & $^{12}$CO & $^{13}$CO & $^{13}$CO\\
   VA procedure & no & yes & no & yes\\
  \hline
  \hline
Center &  &  &  &  &  \\
$I_{\rm CO}$ (K km s$^{-1}$)& $70.0\pm2.0$ (36)\footnotemark[$\ast$] & $70.6\pm2.2$ (32) & $3.37\pm0.53$ (6) & $3.59\pm0.38$ (10) \\
FWHM\footnotemark[$\ast\ast$] (km s$^{-1}$) & $225\pm11$ & $194\pm6$ & $209\pm36$ & $195\pm21$ \\
$T_{\rm peak}$ (mK) & $308\pm20$ (15) & $347\pm22$ (15) & $22.3\pm5.3$ (4) & $18.2\pm3.9$ (5) \\
  \hline
Bar &  &  &  &  &  \\
$I_{\rm CO}$ (K km s$^{-1}$)& $45.1\pm1.6$ (29) & $45.2\pm1.6$ (28) & $1.17\pm0.24$ (5) & $1.42\pm0.09$ (16)\\
FWHM (km s$^{-1}$) & $287\pm34$ & $147\pm7$ & -- & $126\pm14$ \\
$T_{\rm peak}$ (mK) & $241\pm17$ (14) & $323\pm19$ (17) & $8.28\pm2.64$ (3) & $13.4\pm1.1$ (12) \\
  \hline
Bar-end &  &  &  &  &  \\
$I_{\rm CO}$ (K km s$^{-1}$)& $45.5\pm1.3$ (36) & $46.6\pm0.6$ (74) & $3.54\pm0.22$ (16) & $3.56\pm0.11$ (33)\\
FWHM (km s$^{-1}$) & $87\pm5$, $126\pm14$\footnotemark[$\ast\ast\ast$] & $110\pm4$ & $112\pm9$, $76\pm14$ & $100\pm4$ \\
$T_{\rm peak}$ (mK) & $220\pm14$ (15), $215\pm14$ (15) & $400\pm9$ (44) & $18.1\pm2.5$ (7), $20.8\pm2.5$ (8) & $34.5\pm1.6$ (22) \\
  \hline
Offset &  &  &  &  &  \\
$I_{\rm CO}$ (K km s$^{-1}$)& $20.6\pm1.1$ (19) & $20.1\pm0.9$ (23) & $1.46\pm0.25$ (6) & $1.52\pm0.23$ (7)\\
FWHM (km s$^{-1}$) & $67\pm5$ & $67\pm3$ & $50\pm6$ & $54\pm7$ \\
$T_{\rm peak}$ (mK) & $276\pm20$ (14) & $279\pm16$ (17) & $28.7\pm4.5$ (6) & $26.9\pm4.4$ (6) \\
  \hline
Arm &  &  &  &  &  \\
$I_{\rm CO}$ (K km s$^{-1}$) & $29.5\pm1.0$ (31) & $28.6\pm0.6$ (51) & $2.24\pm0.27$ (8) & $2.01\pm0.11$ (19)\\
FWHM (km s$^{-1}$) & $206\pm11$ & $104\pm4$ & $168\pm16$ & $97\pm8$ \\
$T_{\rm peak}$ (mK) & $133\pm11$ (12) & $248\pm7$ (36) & $13.7\pm3.2$ (4) & $19.1\pm1.3$ (15) \\
  \hline
Interarm &  &  &  &  &  \\
$I_{\rm CO}$ (K km s$^{-1}$) & $22.4\pm0.9$ (25) & $20.4\pm1.0$ (20) & $1.02\pm0.27$ (4) & $0.824\pm0.118$ (7)\\
FWHM (km s$^{-1}$) & $268\pm23$ & $159\pm7$ & $170\pm55$ & $94\pm13$ \\
$T_{\rm peak}$ (mK) & $88\pm10$ (9) & $130\pm13$ (10) & $6.89\pm2.82$ (2) & $8.24\pm1.52$ (5) \\
  \hline
\end{tabular}
\end{center}
\footnotetext[$\ast$]{The signal-to-noise ratios of $I_{\rm CO}$ and $T_{\rm peak}$ are shown in the round bracket after each value.}
\footnotetext[$\ast\ast$]{The FWHM is estimated with Gaussian fitting.}
\footnotetext[$\ast\ast\ast$]{ For the stacking results without the VA procedure in the bar-end region, the FWHM and $T_{\rm peak}$ values of the two velocity components in the spectra are shown separately.}
\end{minipage}
\end{table*}

The stacked spectra with and without the VA procedure of $^{12}$CO and $^{13}$CO are shown in figure \ref{fig:StackedSpectra}.
The $^{12}\rm{CO}$ spectra are binned so that the velocity resolution matches that of the $^{13}\rm{CO}$ spectra (i.e., 20 km s$^{-1}$).
The smoothed $^{12}$CO spectra are employed for the following analysis.
The r.m.s. noise temperatures of the stacked $^{12}$CO and $^{13}$CO spectra per $20$ km s$^{-1}$
are typically reduced to $7-22$ mK and $1.1-4.4$ mK, respectively.
The error of the $T_{\rm peak}$ listed in table \ref{tab-2} is the r.m.s. noise temperature, which is calculated within a narrow velocity range outside the emission line.
This is because we must calculate the r.m.s. noise temperature within the baseline range in which the same number of data are stacked as the emission line range.
The number of the stacked data is getting small as the velocity offset from the line center is large.
In table \ref{tab-2}, some errors of $T_{\rm peak}$ and $I_{\rm CO}$ of the stacked spectra with VA are slightly larger than those of the stacked spectra without VA.
This may be partly attributed to the poor statistics in estimation of the r.m.s. noise temperature.

The increase of the peak temperature and the reduction of noise level by the stacking analysis after the VA procedure allow us to detect $^{13}$CO emission even from the interarm region with S/N of 5 where the emission was not detected in the spectrum of each pixel in the original data or in the stacked spectrum without the VA procedure (S/N $=2$).
The fitting results with a Gaussian for the stacked spectra after the VA procedure are also plotted with red lines on each spectrum in figure \ref{fig:StackedSpectra}.
The integrated intensity ($I_{\rm CO}$), full width at half maximum (FWHM) and the peak temperature ($T_{\rm peak}$) of $^{12}$CO and $^{13}$CO with and without the VA procedure in different regions of NGC~3627 are shown in table \ref{tab-2}.
Although the line profiles of the stacked spectra, especially the one without the VA procedure, do not look like a Gaussian, the FWHM and its error are determined by a Gaussian fitting and the fitting error, respectively.
If the FWHM is calculated literally as the width at the half of the peak intensity, the derived FWHM is more susceptible to the r.m.s. noise temperature than the one from global fitting.
In table \ref{tab-2}, we also separately present the FWHM and $T_{\rm peak}$ values of the two velocity components in the bar-end spectrum obtained without the VA procedure.

The validity of this method is examined by comparing the integrated intensity of the stacked spectra with and without the VA procedure.
Both spectra obtained with and without the VA procedure are the averaged spectra in each region.
Therefore, the integrated intensity estimated in both ways should be the same.
In table \ref{tab-2}, we see that the two $I_{^{12}\rm{CO}}$ values of each region are almost the same within the error, confirming the validity of the stacking method with the VA procedure.
Moreover the S/N of $I_{\rm CO}$ of each spectrum is improved by a factor of up to 3.2.

\section{Analyses}\label{Analyses}

\subsection{Surface density of molecular gas mass}

We estimate the surface density of molecular gas mass in the six regions of NGC~3627 from the $^{12}$CO and $^{13}$CO spectra.
$I_{\rm ^{12}CO}$ and $I_{\rm ^{13}CO}$ estimated from Gaussian fitting are used for the calculation hereafter, since the line profiles of the stacked spectra after the VA procedure are well fitted with a Gaussian (see figure \ref{fig:StackedSpectra})\footnote{
The $^{13}$CO emission line of the stacked spectra in the center region has a bluewards wing that is not seen in $^{12}$CO spectra with higher S/N than $^{13}$CO.
Therefore the wing is not expected to be a realistic feature but a wrong feature due to bad baselines of $^{13}$CO spectra.
The Gaussian fitting is not expected to be severely affected by this feature, because the FWHM of $^{13}$CO is consistent with the one of $^{12}$CO within the margin of error.
}.
The integrated intensities estimated with a Gaussian fitting agree with the values listed in table \ref{tab-2} within the margin of error.
The column density of H$_2$, $N_{\rm H_2}$ (cm$^{-2}$), for extragalactic objects is commonly estimated from $I_{\rm ^{12}CO}$ with a CO-to-H$_2$ conversion factor, $X_{\rm CO}$ (cm$^{-2}$ [K km $\rm s^{-1}$]$^{-1}$) under the premise that $^{12}$CO is optically thick.
If $^{12}$CO is optically thick, the brightness temperature is not mainly related to the column density of gas but to the excitation temperature of the $\tau_{\rm ^{12}CO}\sim1$ surface of the virialized molecular clouds \citep{Bolatto+2013}.
$N_{\rm H_2}$ is described as
\begin{equation}
N_{\rm H_2} = X_{\rm CO} I_{\rm ^{12}CO}.
\end{equation}

We can also estimate $N_{\rm H_2}$ from the column density of $^{13}$CO, $N_{\rm ^{13}CO}$ as long as the $^{13}$CO emission is optically thin.
Under the local thermal equilibrium (LTE) approximation, $N_{\rm ^{13}CO}$ can be calculated as
\begin{equation}
N_{\rm ^{13}CO} = \frac{3 k_{\rm B}}{4 \pi^3 \mu^2 \nu_{\rm ^{13}CO}}\exp{\left(-\frac{h\nu_{\rm ^{13}CO}J}{2 k_{\rm B} T_{\rm k}} \right)}\frac{I_{\rm ^{13}CO}}{1-\exp{\left(-\frac{h \nu_{\rm ^{13}CO}}{k_{\rm B} T_{\rm k}} \right)}}\ \ \ \rm{cm^{-2}},
\label{eq-4}
\end{equation}
where $k_{\rm B}$ is the Boltzmann constant, $\mu$ is the dipole moment of $0.11\times10^{-18}$ esu cm, $h$ is the Planck constant, $\nu_{\rm ^{13}CO}$ is the rest-frame frequency of $^{13}$CO, $J$ is the rotational quantum number of lower energy state and $T_{\rm k}$ is the kinetic temperature.
We obtain $(N_{\rm H_2}/{\rm cm}^{-2})=8.07\times10^{20} (I_{\rm ^{13}CO}/\rm{K\ km\ s}^{-1})$ by assuming an $N_{\rm H_2}/N_{\rm ^{13}CO}$ ratio of $7.5\times10^5$ \citep{Frerking+1982} and $T_{\rm k}=20$ K.

It is useful to estimate a lower limit of $N_{\rm H_2}$ by assuming that the $^{12}$CO line has a small optical depth ($\tau_{\rm ^{12}CO} \ll 1.0$).
The minimum H$_2$ column density is calculated in the same manner as equation (\ref{eq-4}) with $I_{\rm ^{12}CO}$.
We obtain $(N_{\rm H_2}/{\rm cm^{-2}})=9.88\times10^{18} (I_{\rm ^{12}CO}/\rm{K\ km\ s^{-1}})$ with $N_{\rm H_2}/N_{\rm ^{12}CO}$ ratio of $1.0\times10^4$ \citep{YoungScoville1991} and $T_{\rm k}=20$ K.
We utilize this conversion factor in the discussion in section \ref{Discussion}.

The surface density of the molecular gas, $\Sigma_{\rm mol}$ is calculated as,
\begin{equation}
\Sigma_{\rm mol}=1.36\times 2 \times m_{\rm H} N_{\rm H_2} \cos{(i)},
\end{equation}
where 1.36 is a factor to account for the contribution of He by mass, $m_{\rm H}$ is the mass of the hydrogen atom and $i$ is the inclination of NGC~3627 ($52^\circ$).
Then we can obtain the surface density of the molecular gas for the case of optically thick $^{12}$CO emission as,
\begin{equation}
\left( \frac{\Sigma_{\rm mol,12,thick}}{M_\odot\ \rm{pc^{-2}}} \right ) = 1.34\ \left( \frac{I_{\rm ^{12}CO}}{\rm K\ km\ s^{-1}} \right),
\label{eq:mol12thick}
\end{equation}
for the case of optically thin $^{13}$CO emission as,
\begin{equation}
\left( \frac{\Sigma_{\rm mol,13,thin}}{M_\odot\ \rm{pc^{-2}}} \right ) = 10.8\ \left( \frac{I_{\rm ^{13}CO}}{\rm K\ km\ s^{-1}} \right),
\label{eq:mol13thin}
\end{equation}
and for the case of optically thin $^{12}$CO emission as,
\begin{equation}
\left( \frac{\Sigma_{\rm mol,12,thin}}{M_\odot\ \rm{pc^{-2}}} \right ) = 0.133\ \left( \frac{I_{\rm ^{12}CO}}{\rm K\ km\ s^{-1}} \right),
\label{eq:mol12thin}
\end{equation}
where $\Sigma_{\rm mol,12,thick}$, $\Sigma_{\rm mol,13,thin}$ and $\Sigma_{\rm mol,12,thin}$ are the surface densities of molecular gas estimated from $^{12}$CO (optically thick), $^{13}$CO (optically thin) and $^{12}$CO (optically thin), respectively.
We adopt $X_{\rm CO}$ of $1\times10^{20}$ cm$^{-2}$ $[\rm K\ km\ s^{-1}]^{-1}$ \citep{NakaiKuno1995}.
The $\Sigma_{\rm mol,12,thick}$, $\Sigma_{\rm mol,13,thin}$, $\Sigma_{\rm mol,12,thin}$ and $\Sigma_{\rm mol,12,thick}/\Sigma_{\rm mol,13,thin}$ ratios in all the regions are shown in table \ref{tab-3}.

In table \ref{tab-3}, there is a wide variety in $\Sigma_{\rm mol,12,thick}/\Sigma_{\rm mol,13,thin}$ ratios among the six regions.
Here we focus on the relative difference of the $\Sigma_{\rm mol,12,thick}/\Sigma_{\rm mol,13,thin}$ ratios among different regions rather than the discrepancy between $\Sigma_{\rm mol,12,thick}$ and $\Sigma_{\rm mol,13,thin}$
in each region.
The errors of $\Sigma_{\rm mol}$ in table \ref{tab-3} are calculated only from the errors of the integrated intensities.
However, the $\Sigma_{\rm mol}$ values calculated here implicitly include the following assumptions;
1) the observed molecular clouds mainly consists of H$_2$ molecules, are virialized, follow the size-line width relation and have constant temperature and the $^{12}$CO emission from them is optically thick (for the constant CO-to-H$_2$ conversion factor),
2) the kinetic temperature and the abundance ratios of $^{13}$CO to H$_2$ are free parameters (for the LTE assumption).
In particular, the coefficient in equation (\ref{eq:mol13thin}) varies from 6.16 to 25.1 for $T_{\rm k}=10$ and $50$ K, respectively.

We estimate $T_{\rm k}$ from the comparison between $\Sigma_{\rm mol, 12, thick}$ and $\Sigma_{\rm mol, 13, thin}$ by assuming that both $^{12}$CO and $^{13}$CO emissions are radiated from the same region, the abundance ratios of $^{13}$CO to H$_2$ do not vary within the galaxy, and the estimated value of $\Sigma_{\rm mol, 12, thick}$ with $X_{\rm CO}$ is correct.
In table \ref{tab-3}, we can see that the $\Sigma_{\rm mol, 12, thick}/\Sigma_{\rm mol, 13, thin}$ ratios in the center, bar and interarm regions are $\sim3$ while those in the other regions are as small as $\sim1.6$.
We obtain $T_{\rm k}\sim50$ K for the center, $\sim90$ K for the bar, $\sim70$ K for the interarm, and $\sim30-40$ K for the other regions.
\citet{Galametz+2012} derived the dust temperature distribution of NGC~3627 with a grid size of $18''$, which corresponds to half of the effective spatial resolution of $36''$.
They found a radial temperature gradient declining from $\sim25$ K to $\sim17$ K from their SED fitting using the dust temperature and emissivity index as free parameters (figure 4 of \cite{Galametz+2012}).
In their plot, the highest temperature ($\sim25$ K) is found in the center and the bar-end regions and the lowest values are seen in the interarm region ($\sim17$ K).
In the center region, which contains nuclear starburst and active galactic nuclei \citep{Krips+2008}, the high $\Sigma_{\rm mol, 12, thick}/\Sigma_{\rm mol, 13, thin}$ ratio may be partly attributed to the high $T_{\rm k}$ although the averaged temperature of $\sim50$ K over the 800-pc beam is too high.
However, it is quite unlikely that the temperatures of the molecular gas in the bar and interarm regions are higher than the arm, bar-end and offset regions where stars are actively forming.
Therefore, the high $\Sigma_{\rm mol, 12, thick}/\Sigma_{\rm mol, 13, thin}$ ratios in the bar and the interarm regions are likely attributed to the overestimation of $\Sigma_{\rm mol, 12, thick}$ estimated with $I_{\rm ^{12}CO}$ and a constant CO-to-H$_2$ conversion factor.
We compare the $^{12}$CO and $^{13}$CO spectra of each region in the following subsections to physically explain the high $\Sigma_{\rm mol, 12, thick}/\Sigma_{\rm mol, 13, thin}$ ratios in the bar and interarm regions.

\begin{table*}
 \caption{The surface densities of molecular gas $\Sigma_{\rm mol}$ estimated under the assumption of optically thick $^{12}$CO (equation (\ref{eq:mol12thick})), optically thin $^{13}$CO (equation (\ref{eq:mol13thin})) and optically thin $^{12}$CO (equation (\ref{eq:mol12thin})).}
 \label{tab-3}
 \begin{minipage}{\textwidth}
 \begin{center}
 \begin{tabular}{@{}lcccccc}
  \hline
 & Center & Bar & Bar-end & Offset & Arm & Interarm\\
  \hline
  \hline
$\Sigma_{\rm mol,12,thick}$ ($M_\odot$ pc$^{-2})$& $93.7\pm3.7$ & $60.2\pm3.7$ & $63.4\pm3.0$ & $27.6\pm1.8$ & $36.0\pm1.8$ & $28.1\pm1.6$\\
$\Sigma_{\rm mol,13,thin}$\footnotemark[$\ast$] ($M_\odot$ pc$^{-2}$)& $38.4\pm5.5$ & $17.5\pm2.6$ & $39.5\pm2.3$ & $17.4\pm2.8$ & $20.5\pm2.3$ & $8.32\pm1.51$\\
$\Sigma_{\rm mol,12,thin}$\footnotemark[$\ast$] ($M_\odot$ pc$^{-2}$)& $9.30\pm0.37$ & $5.98\pm0.36$ & $6.29\pm0.30$ & $2.74\pm0.18$ & $3.58\pm0.18$ & $2.79\pm0.15$ \\
$\Sigma_{\rm mol,12,thick}/\Sigma_{\rm mol,13,thin}$\footnotemark[$\ast\ast$] & $2.4\pm0.4$ & $3.4\pm0.6$ & $1.6\pm0.1$ & $1.6\pm0.3$ & $1.8\pm0.2$ & $3.0\pm0.6$\\
  \hline
  \end{tabular}
\end{center}
\footnotetext[$\ast$]{$\Sigma_{\rm mol}$ is estimated with $T_{\rm k}=20$.}
\footnotetext[$\ast\ast$]{$\Sigma_{\rm mol,12,thick}/\Sigma_{\rm mol,13,thin}$ ratio is also shown.}
\end{minipage}
\end{table*}

\subsection{Integrated intensity, FWHM, and peak temperature ratios of the $^{12}$CO and $^{13}$CO spectra}\label{subsec-Discussion2}

We show the $I_{^{12}\rm{CO}}$/$I_{^{13}\rm{CO}}$, FWHM$_{^{12}\rm{CO}}$/FWHM$_{^{13}\rm{CO}}$ and $T_{\rm peak, ^{12}CO}/T_{\rm peak, ^{13}CO}$ ratios in table \ref{tab-4}.
We find for the first time that the $I_{^{12}\rm{CO}}$/$I_{^{13}\rm{CO}}$ ratio in the interarm region is almost twice as high as those in the bar-end, offset and arm regions.
The high ratios in the bar and center regions reported in W11\nocite{Watanabe+2011} are confirmed by the stacking analysis.
The $I_{^{12}\rm{CO}}$/$I_{^{13}\rm{CO}}$ ratios obtained in the bar-end, offset and arm regions are consistent with the values obtained in \citet{Paglione+2001} where they observed 17 nearby galaxies in $^{12}$CO and $^{13}$CO along the major axes and obtained the $I_{^{12}\rm{CO}}/I_{^{13}\rm{CO}}$ ratios of $4-22.8$ ($45''$ spatial resolution).
The bar region shows a higher value of $T_{\rm peak, ^{12}CO}/T_{\rm peak, ^{13}CO}=24.0$ than the other regions ($10-16$) and the center region shows an intermediate value between them ($19.1$).

It is noteworthy that the FWHM$_{^{12}\rm{CO}}$/FWHM$_{^{13}\rm{CO}}$ ratio in the interarm region is $1.7$ whereas the other regions have almost unity.
In other words, the line width of $^{12}$CO is larger than that of $^{13}$CO in the interarm region.
To investigate local effects on this trend, we separate the interarm region into Northeastern (NE) and Southwestern (SW) parts, produce the averaged spectra in each part, measure the FWHM ratios, and check whether this trend still holds or not. 
Their FWHM estimated with Gaussian fitting are presented in table \ref{tab-5}.
Although the S/N (the peak temperature-to-noise ratio) of the data is not so high ($\sim4$), the trend that FWHM of $^{12}$CO is larger than that of $^{13}$CO still exists.
The FWHM ratios that are estimated with NE and SW interarm spectra separately are consistent with the interarm ratio of $\sim1.7$ within the margin of error.
Thus we conclude that the difference in FWHM between the $^{12}$CO and $^{13}$CO spectra is a characteristic feature in the interarm region of NGC~3627 rather than any local feature, which has nothing to do with the galactic structures.

\begin{table*}
 \caption{The $^{12}$CO$/$$^{13}$CO ratios of intensity, FWHM and $T_{\rm peak}$ estimated from Gaussian fitting in each region.}
 \label{tab-4}
 \begin{tabular}{@{}lcccccc}
  \hline
   & Center & Bar & Bar-end & Offset & Arm & Interarm\\
  \hline
  \hline
$I_{^{12}\rm{CO}}$/$I_{^{13}\rm{CO}}$ & $19.7\pm1.4$ & $31.8\pm3.9$ & $13.1\pm0.5$ & $13.3\pm1.4$ & $14.2\pm0.6$ & $24.7\pm2.8$\\
FWHM$_{^{12}\rm{CO}}$/FWHM$_{^{13}\rm{CO}}$ & $0.99\pm0.11$ & $1.17\pm0.14$ & $1.10\pm0.06$ & $1.24\pm0.17$ & $1.07\pm0.10$ & $1.69\pm0.24$\\
$T_{\rm peak, ^{12}CO}/T_{\rm peak, ^{13}CO}$ & $19.1\pm2.8$ & $24.0\pm3.8$ & $11.6\pm0.7$ & $10.4\pm1.0$ & $13.0\pm0.7$ & $15.8\pm2.4$ \\
  \hline
  \end{tabular}
\end{table*}

\begin{table*}
\caption{FWHM estimated with Gaussian fitting to the stacked spectra of the interarm region.}
\begin{center}
\begin{tabular}{lccc}
\hline
&FWHM$_{^{12}\rm{CO}}$&FWHM$_{^{13}\rm{CO}}$&FWHM$_{^{12}\rm{CO}}$/FWHM$_{^{13}\rm{CO}}$\\
& (km s$^{-1}$)&(km s$^{-1}$)&\\
\hline
\hline
All & $159\pm7$ & $94\pm13$ & $1.69\pm0.24$\\
Northern-East & $145\pm9$ & $66\pm12$ & $2.20\pm0.43$\\
Southern-West & $175\pm8$ & $123\pm16$ & $1.42\pm0.19$\\
\hline
\end{tabular}
\end{center}
\label{tab-5}
\end{table*}%

\subsubsection{The radial trends of $I_{\rm ^{12}CO}$/$I_{\rm ^{13}CO}$, FWHM$_{\rm ^{12}CO}$/FWHM$_{\rm ^{13}CO}$ and $T_{\rm peak, ^{12}CO}$/$T_{\rm peak, ^{13}CO}$}

\begin{figure*}
\includegraphics[width=170mm]{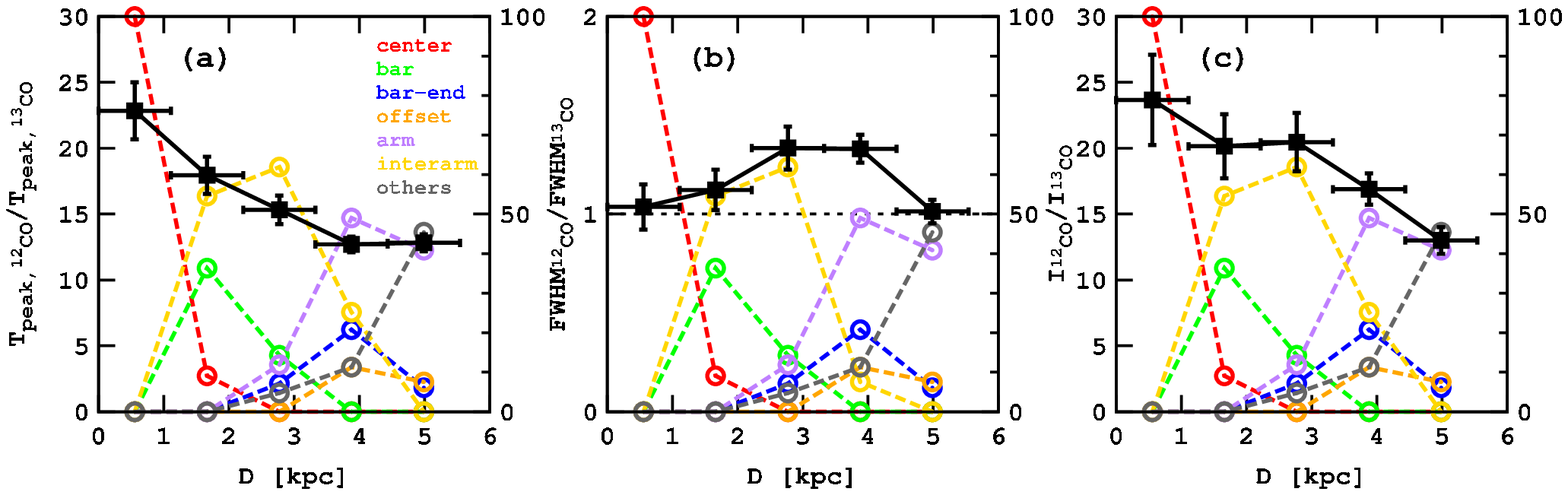}
 \vspace{0cm}
 \caption{Radial distributions of a) $T_{\rm peak, ^{12}CO}/T_{\rm peak, ^{13}CO}$, b) FWHM$_{^{12}\rm{CO}}$/FWHM$_{^{13}\rm{CO}}$ and c) $I_{^{12}\rm{CO}}$/$I_{^{13}\rm{CO}}$.
 These ratios are shown as a black line in each graph.
 Red, green, blue, orange, purple, yellow, and grey lines represent the fraction of spectra of center, bar, bar-end, arm, interarm, offset regions and others included in each concentric annulus.
 The error bars of the three ratios (vertical axis) are estimated from the fitting error (1$\sigma$).
The horizontal bar of each point represents the radial range within which the stacked spectrum is calculated.
 }
  \label{fig:RadialDistributions}
\end{figure*}

Each region has a different mean galactocentric distance and therefore the variations in $I_{\rm ^{12}CO}$/$I_{\rm ^{13}CO}$, FWHM$_{\rm ^{12}CO}$/FWHM$_{\rm ^{13}CO}$ and $T_{\rm peak, ^{12}CO}$/$T_{\rm peak, ^{13}CO}$ found in the previous section may be attributed to a radial trend.
To investigate the radial gradient of $I_{\rm ^{12}CO}$/$I_{\rm ^{13}CO}$, FWHM$_{\rm ^{12}CO}$/FWHM$_{\rm ^{13}CO}$ and $T_{\rm peak, ^{12}CO}$/$T_{\rm peak, ^{13}CO}$, we produced the stacked spectra of $^{12}$CO and $^{13}$CO in five concentric annuli (r1$-$r5, from the galaxy center) with a width of $\sim1.1$ kpc in the galactic plane.
The fitting results with a Gaussian to the stacked spectra after the VA procedure of the five annuli are summarized in table \ref{tab-6}.
The radial profiles of the $I_{\rm ^{12}CO}$/$I_{\rm ^{13}CO}$, FWHM$_{\rm ^{12}CO}$/FWHM$_{\rm ^{13}CO}$ and $T_{\rm peak, ^{12}CO}$/$T_{\rm peak, ^{13}CO}$ ratios are shown in figure \ref{fig:RadialDistributions} with black lines.
In these plots, we also show the fraction of the number of pixels of each six morphologically defined regions included in a concentric annulus to the total number of pixels in each annulus.
The colors of these lines are the same as the six regions in figure \ref{fig:ProfileMap}.
Grey points and lines (``others'') represent the area which is not categorized in the (1)$-$(6) regions but included in the concentric annuli.

We find a radial gradient in the $T_{\rm peak, ^{12}CO}/T_{\rm peak, ^{13}CO}$ ratio plot (figure \ref{fig:RadialDistributions}a) and a weak gradient in the $I_{^{12}\rm{CO}}$/$I_{^{13}\rm{CO}}$ ratio plot (figure \ref{fig:RadialDistributions}c).
The radial gradient of $I_{^{12}\rm{CO}}$/$I_{^{13}\rm{CO}}$ reported in W11\nocite{Watanabe+2011} is confirmed with the data obtained with the stacking analysis.
The $T_{\rm peak, ^{12}CO}/T_{\rm peak, ^{13}CO}$ and the $I_{^{12}\rm{CO}}$/$I_{^{13}\rm{CO}}$ ratios tend to be higher at smaller galactocentric distance $D$ (kpc) while we found the highest values of both ratios in the bar and interarm regions outside r1.
In figure \ref{fig:RadialDistributions}b, we do not see any systematic radial gradient in the FWHM$_{^{12}\rm{CO}}$/FWHM$_{^{13}\rm{CO}}$ plot but there is a bump at r2$-$r4.
This radial range contains the interarm region and the highest contribution from the interarm region is found in r2$-$r3.
Additionally, the FWHM$_{^{12}\rm{CO}}$/FWHM$_{^{13}\rm{CO}}$ ratios at the bump ($\sim1.3$) are smaller than those of the interarm region ($\sim1.7$).
Hence, we conclude that the high FWHM$_{\rm ^{12}CO}$/FWHM$_{\rm ^{13}CO}$ ratios at r2$-$r4 are due to the inclusion of the interarm region.
Accordingly, the differences seen in the $I_{\rm ^{12}CO}$/$I_{\rm ^{13}CO}$, FWHM$_{\rm ^{12}CO}$/FWHM$_{\rm ^{13}CO}$ and $T_{\rm peak, ^{12}CO}$/$T_{\rm peak, ^{13}CO}$ ratios in the morphologically defined regions do not represent the radial trends of these ratios.

\begin{table*}
\caption{Gaussian fitting results of the stacked spectra after the VA procedure in different concentric radii.}
\begin{center}
\begin{tabular}{lccc}
\hline
Line & $^{12}$CO & $^{13}$CO & $^{12}$CO/$^{13}$CO\\
\hline
\hline
r1: $D=0-1.1$ (kpc)&&&\\
$I_{\rm CO}$ (K km s$^{-1}$) & $72.7\pm2.8$ & $3.07\pm43$ & $23.7\pm3.4$\\
FWHM (km s$^{-1}$) & $191\pm5$ & $184\pm20$ & $1.03\pm0.11$\\
$T_{\rm peak}$ (mK) & $357\pm9$ & $15.6\pm1.4$ & $22.8\pm2.2$\\
\hline
r2: $1.1-2.2$ (kpc)&&&\\
$I_{\rm CO}$ (K km s$^{-1}$) & $35.4\pm1.3$ & $1.76\pm0.20$ & $20.1\pm3.4$\\
FWHM (km s$^{-1}$) & $167\pm5$ & $149\pm13$ & $1.12\pm0.10$\\
$T_{\rm peak}$ (mK) & $199\pm5$ & $11.1\pm0.8$ & $17.9\pm1.4$\\
\hline
r3: $2.2-3.3$ (kpc)&&&\\
$I_{\rm CO}$ (K km s$^{-1}$) & $25.7\pm1.5$ & $1.25\pm0.11$ & $20.5\pm2.2$\\
FWHM (km s$^{-1}$) & $144\pm6$ & $108\pm7$ & $1.33\pm0.11$\\
$T_{\rm peak}$ (mK) & $167\pm6$ & $10.9\pm0.7$ & $15.3\pm1.1$\\
\hline
r4: $3.3-4.4$ (kpc)&&&\\
$I_{\rm CO}$ (K km s$^{-1}$) & $30.4\pm1.2$ & $1.80\pm0.11$ & $16.9\pm1.2$\\
FWHM (km s$^{-1}$) & $94.1\pm2.8$ & $70.7\pm3.1$ & $1.33\pm0.1$\\
$T_{\rm peak}$ (mK) & $304\pm8$ & $23.9\pm0.9$ & $12.7\pm0.6$\\
\hline
r5: $4.4-5.5$ (kpc)&&&\\
$I_{\rm CO}$ (K km s$^{-1}$) & $20.2\pm1.0$ & $1.56\pm0.09$ & $13.0\pm1.0$\\
FWHM (km s$^{-1}$) & $119\pm4$ & $118\pm5$ & $1.01\pm0.06$\\
$T_{\rm peak}$ (mK) & $160\pm5$ & $12.5\pm0.5$ & $12.8\pm0.7$\\
\hline
\end{tabular}
\end{center}
\label{tab-6}
\end{table*}%

\subsection{The intensity ratios of $T_{^{12}\rm{CO}}/T_{^{13}\rm{CO}}$ in the different regions}
\label{subsec:slope}

\begin{figure*}
\includegraphics[width=110mm]{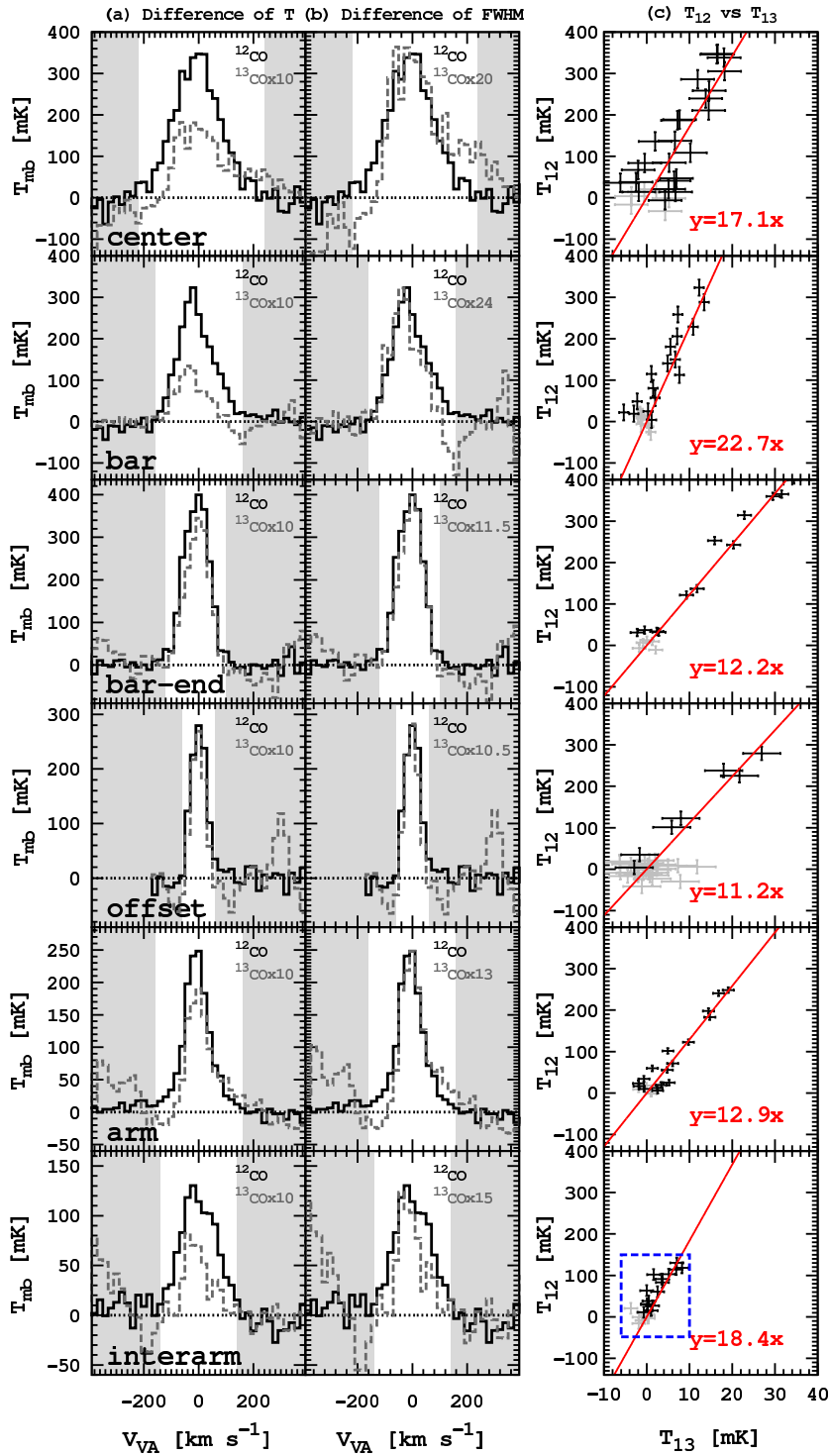}
\vspace{+1.0cm}
 \caption{Comparison between the $^{12}$CO and $^{13}$CO spectra.
(a) $^{12}$CO and $10\times ^{13}$CO spectra for a comparison of $T_{\rm mb}$.
(b) $^{12}$CO and $^{13}$CO spectra multiplied by adequate values where the peak temperatures of $^{12}$CO and $^{13}$CO become comparable for a comparison of FWHM.
The ``emission'' range of spectra is indicated as an unshaded region and the ``baseline'' ranges are indicated as shaded regions.
(c) $T_{\rm ^{12}CO}-T_{\rm ^{13}CO}$ correlation plots.
Grey and black points in the plot represent the data points of ``baseline'' and ``emission'' range of the spectra.
The linear fitting results of data within the ``emission'' range (black points) are also shown as a black solid line.
The error bars represent the r.m.s. noise temperatures of the $^{12}$CO and $^{13}$CO spectra ($1\sigma$).
They are fitted so as to cross the origin.
A magnified figure of the dotted rectangle region on the plot of the interarm region is shown in figures \ref{fig:SpectraComparison1213Interarm}b and \ref{fig:SpectraComparison1213Interarm}c.
}
  \label{fig:SpectraComparison1213}
\end{figure*}

We compare the $^{12}\rm{CO}$ and $^{13}\rm{CO}$ spectra in each region in figure \ref{fig:SpectraComparison1213}.
Column (a) of this figure is $^{12}$CO and $^{13}$CO spectra multiplied by a fixed value of $10$ for the comparison of the heights of the $^{12}$CO and the $^{13}$CO emission lines in each different region.
Column (b) shows the $^{12}$CO and the rescaled $^{13}$CO spectra, which are matched at the peak of the $^{12}$CO.
This is done to compare the widths of the $^{12}$CO and the $^{13}$CO emission lines.
The emission ranges of the spectra are indicated as the unshaded regions and the baseline range are indicated as the shaded regions in the column (a) and (b).
Hereafter we refer to them as the ``emission'' and ``baseline'' ranges, respectively.
In column (c) of this figure, we plot the $T_{\rm mb}$ of $^{12}$CO versus $^{13}$CO for all the velocity channels in each region.
Grey and black points in this plot represent the data points of the ``baseline'' and ``emission'' ranges of the spectra.
The linear fitting results of the data within the ``emission'' range (black-filled circles) are also shown in red solid lines.
They are fitted so as to cross the origin.

Column (a) in figure \ref{fig:SpectraComparison1213} visually confirms that the $^{12}$CO spectra of the center, bar and interarm regions are still much higher than $10\times^{13}$CO, unlike those in the three other regions.
This difference is quantitatively expressed by different slopes of the fitting lines in column (c) in figure \ref{fig:SpectraComparison1213}.
The slope of 22.7 in the bar region is almost twice as high as those in the bar-end, offset and arm regions ($11-13$) and those in the center and interarm regions show the intermediate value ($17-18$) between them.

\begin{figure*}
\includegraphics[width=150mm]{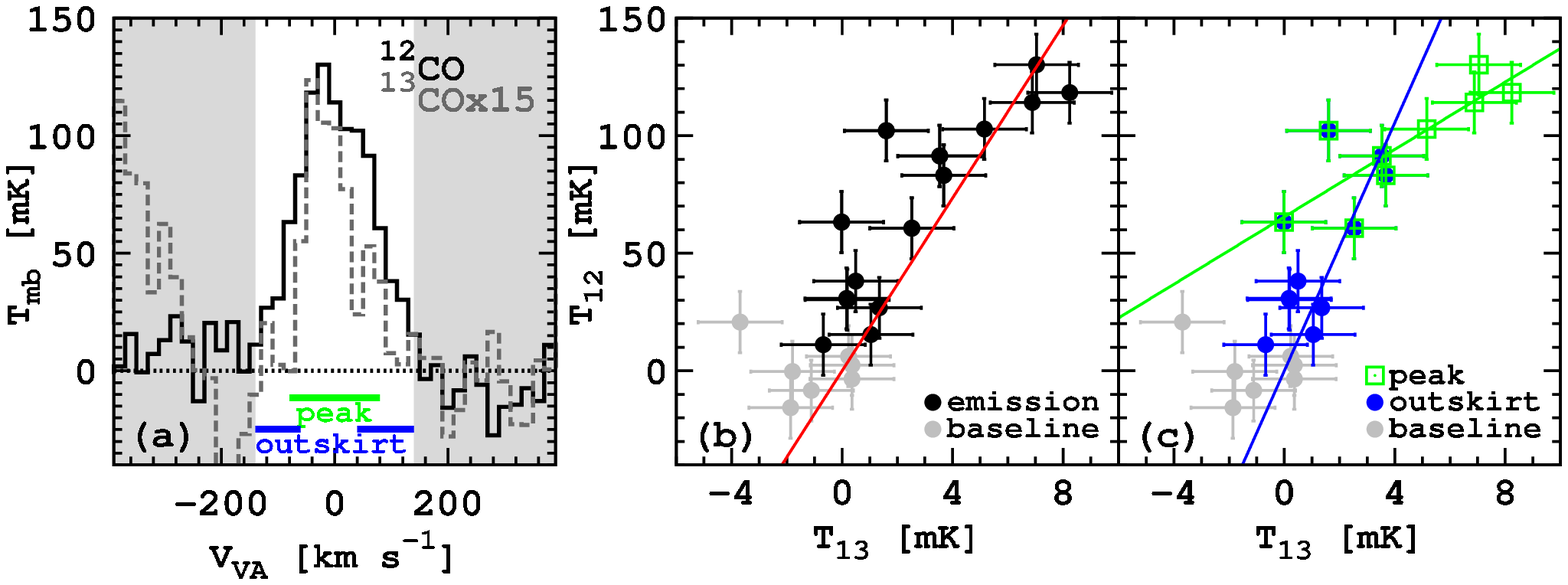}
 \caption{Comparison between the $^{12}$CO and $^{13}$CO spectra in the interarm region.
(a) $^{12}$CO and $^{13}$CO spectra multiplied by 15.
The ``baseline,'' ``peak,'' and ``outskirt'' ranges are indicated as shaded region, green and blue horizontal lines, respectively.
(b) A magnified figure of the dotted rectangle region in the original plot in figure \ref{fig:SpectraComparison1213}c.
(c) The same plot as (b).
But the data points are shown in different colors and symbols:
grey filled circles for ``baseline,'' blue filled circles for ``outskirt,'' and green open square for ``peak.''
The blue and green solid lines on this plot represent the fitting results of the data within the ``outskirt'' and ``peak'' ranges, respectively.
The error bars in (b) and (c) represent the r.m.s. noise temperatures of the $^{12}$CO and $^{13}$CO spectra ($1\sigma$).
}
  \label{fig:SpectraComparison1213Interarm}
\end{figure*}

The most notable result in this study is that the width of the $^{12}$CO emission line is larger than that of $^{13}$CO in the interarm region whereas both lines in the other regions have comparable widths (column (b) in \ref{fig:SpectraComparison1213}).
We separate the ``emission'' range of the interarm spectra into the ``peak'' and the ``outskirt'' ranges that are indicated with blue and green horizontal lines in figure \ref{fig:SpectraComparison1213Interarm}a.
In figure \ref{fig:SpectraComparison1213Interarm}b, we show a magnified plot of the dotted rectangle region of figure \ref{fig:SpectraComparison1213}c.
Figure \ref{fig:SpectraComparison1213Interarm}c is the same as figure \ref{fig:SpectraComparison1213Interarm}b, but the data points of the ``peak'' and the ``outskirt'' ranges of the spectra are indicated as blue and green points, respectively.
The blue and green solid lines represent the linear fitting results of the data within the ``outskirt'' and the ``peak'' points.

In figure \ref{fig:SpectraComparison1213Interarm}c, there are distinct spectral features of 1) a large gradient ($26.4\pm5.3$) at $T_{\rm ^{13}CO}$\hspace{0.3em}\raisebox{0.4ex}{$<$}\hspace{-0.75em}\raisebox{-.7ex}{$\sim$}\hspace{0.3em}3 mK and 2) a small gradient ($7.2\pm1.9$) at $T_{\rm ^{13}CO}$\hspace{0.3em}\raisebox{0.4ex}{$>$}\hspace{-0.75em}\raisebox{-.7ex}{$\sim$}\hspace{0.3em}3 mK.
The two slopes are significantly different taking into account the error.
This result suggests that there are two gas components with different velocity widths and the component with a broader line width has the highest $T_{\rm ^{12}CO}/T_{\rm ^{13}CO}$ ratio of $26.4$ among all the regions.

\subsection{Brief summary of the analyses}

The surface densities of molecular gas were calculated with the $^{12}$CO and $^{13}$CO spectra in the six regions of NGC~3627.
We found that the bar and interarm regions have higher $\Sigma_{\rm ^{12}CO, thick}/\Sigma_{\rm ^{13}CO, thin}$, i.e., $I_{\rm ^{12}CO}/I_{\rm ^{13}CO}$ than the other regions.
The $I_{^{12}\rm{CO}}/I_{^{13}\rm{CO}}$ ratio in the bar region is high because the $T_{^{12}\rm{CO}}/T_{^{13}\rm{CO}}$ ratios are fairly high ($22.7$) within the velocity range of the emission line.
For the case of the interarm region, the high $I_{^{12}\rm{CO}}/I_{^{13}\rm{CO}}$ ratio is attributed to the broader line width of the $^{12}$CO spectra compared to the $^{13}$CO spectra.
The difference of the FWHM of $^{12}$CO and $^{13}$CO suggests the existence of two molecular gas components with different FWHM in the interarm region.
$T_{^{12}\rm{CO}}/T_{^{13}\rm{CO}}$ ratio of the gas component with broader FWHM in the interarm region is higher than the other regions.

\section{Discussion}\label{Discussion}

A new and the most important result of our study is that the $I_{\rm ^{12}CO}/I_{\rm ^{13}CO}$ in the interarm region is high due to broader FWHM$_{\rm ^{12}CO}$ than FWHM$_{\rm ^{13}CO}$.
The obtained spectra in this study are the emissions from an ensemble of giant molecular clouds (GMCs), giant molecular associations (GMAs) and/or the ambient components, since the typical sizes of those structures are \hspace{0.3em}\raisebox{0.4ex}{$<$}\hspace{-0.75em}\raisebox{-.7ex}{$\sim$}\hspace{0.3em}$40$ pc (GMCs) and $\sim200$ pc (GMAs), which are much smaller than the beam size of $\sim800$ pc.
Therefore, the observed FWHM of the $^{12}$CO and $^{13}$CO spectra are likely to represent not a random motion within a cloud but a random motion between clouds since the FWHM of the former (\hspace{0.3em}\raisebox{0.4ex}{$<$}\hspace{-0.75em}\raisebox{-.7ex}{$\sim$}\hspace{0.3em}$10$ km s$^{-1}$, \cite{Sanders+1985,Scoville+1987,Solomon+1987}) is much smaller than that of the latter.
If the molecular gas (or clouds) within a beam has almost the same $^{12}$CO to $^{13}$CO intensity ratio in each velocity channel, the FWHM of the spectra of the $^{12}$CO and $^{13}$CO lines are expected to be almost the same.
This  is the case for every region except for the interarm region.
The different FWHM between the $^{12}$CO and $^{13}$CO lines in the interarm region is hardly explained by the ensemble of molecular gas with uniform physical states within a beam.
We discuss the physical conditions of molecular gas in each region of NGC~3627 in the following subsections.

\subsection{Differences in the physical states of molecular gas among different regions}
\label{subsec:physicalstate}

\begin{figure}
\includegraphics[width=70mm]{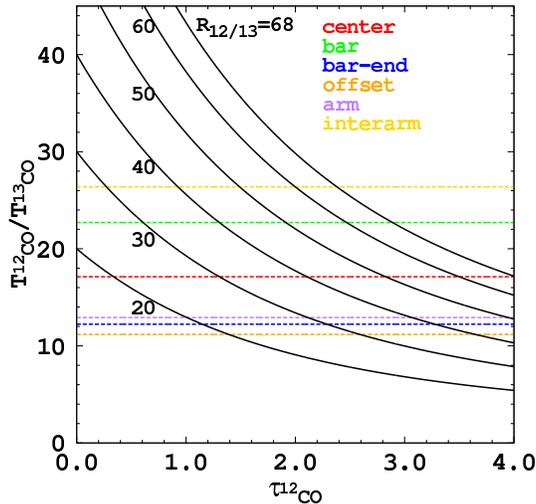}
 \vspace{0cm}
 \caption{Brightness temperature ratio, $\frac{T_{\rm ^{12}CO}}{T_{\rm ^{13}CO}}$ as a function of $\tau_{\rm ^{12}CO}$ in case of $R_{12/13}=68,\ 60,\ 50,\ 40,\ 30,\ 20$.
 The horizontal lines show $\frac{T_{\rm ^{12}CO}}{T_{\rm ^{13}CO}}$ in the center (red), bar (green), bar-end (blue), offset (orange), arm (purple) regions and the ``outskirt'' component of the $^{12}$CO spectrum in the interarm region (yellow).
 }
  \label{fig:tau12CO}
\end{figure}

The brightness temperature of line radiation, $T_{\rm B}(\nu)$, under the assumption of local thermal equilibrium (LTE), is described as
\begin{equation}
T_{\rm{B}}(\nu) = \Phi [ J_\nu (T_{\rm ex}) - J_\nu (T_{\rm bg}) ] [1 - \exp{(-\tau_\nu)}],
\end{equation}
where $\Phi$ is the beam-filling factor, $J_\nu (T)$ is the Plank function $(J_\nu (T) = [2h\nu^3/c^2] [\exp{(h\nu/k_B T)} - 1]^{-1}$, where $c$ is the speed of light), $T_{\rm ex}$ is the excitation temperature and equals the kinetic temperature $T_{\rm k}$ under LTE conditions, $T_{\rm bg}$ is the temperature of the cosmic microwave background ($2.73$ K) and $\tau_\nu$ is the optical depth.
Under the assumptions that 1) both $^{12}$CO and $^{13}$CO lines are emitted from the same cloud (same $\Phi$ for $^{12}$CO and $^{13}$CO), and 2) both molecules are thermalized (same $T_{\rm ex}=T_{\rm k}$ for $^{12}$CO and $^{13}$CO), the line ratio of ${T_{^{12}\rm{CO}}}/{T_{^{13}\rm{CO}}}$ can be described as a function of the optical depth of $^{12}$CO as,
\begin{equation}
\frac{T_{^{12}\rm{CO}}}{T_{^{13}\rm{CO}}} \approx \frac{1-\rm{exp}(-\tau_{^{12}\rm{CO}})}{1-\rm{exp}(-\tau_{^{13}\rm{CO}})} \approx \frac{1-\rm{exp}(-\tau_{^{12}\rm{CO}})}{1-\rm{exp}(-\frac{\tau_{^{12}\rm{CO}}}{\it R_{\rm 12/13}})},
\label{eq:12}
\end{equation}
where $R_{12/13}$ is the abundance ratio of $^{12}$C and $^{13}$C.
$R_{12/13}$ can be used as a proxy for the $N_{\rm ^{12}CO}/N_{\rm ^{13}CO}$ ratio as long as the isotope fractionation is negligible\footnote{
To be precise, the $N_{\rm ^{12}CO}/N_{\rm ^{13}CO}$ ratio may be affected by the isotope fractionation in response to the competing processes of the isotope exchange reaction \citep{Watson+1976,SmithAdams1980} and the selective dissociation \citep{BallyLanger1982,vanDishoeck+1988,Kopp+1996}.
}.
The $T_{^{12}\rm{CO}}/T_{^{13}\rm{CO}}$ ratio as a function of $\tau_{\rm ^{12}CO}$ is shown in figure \ref{fig:tau12CO}.
According to the equation (\ref{eq:12}),
a larger $\tau_{\rm ^{12}CO}$ gives us a lower $T_{^{12}\rm{CO}}/T_{^{13}\rm{CO}}$ with a fixed $R_{12/13}$ and a lower $R_{12/13}$ gives us a lower $T_{^{12}\rm{CO}}/T_{^{13}\rm{CO}}$.
\citet{Milam+2005} showed a radial dependence of $R_{12/13}$ of our Galaxy as
\begin{equation}
R_{12/13}(D) = 6.21\ D +18.71,
\label{eq-R1213}
\end{equation}
where $D$ is the galactocentric distance in kpc.
With an assumption that the profile of the emission line is expressed by a Gaussian function, the optical depth of the $^{13}$CO line can be described as
\begin{equation}
\tau_{\rm ^{12}CO} \approx \frac{4 \pi^3 \nu_{\rm ^{12}CO} \mu^2 N_{\rm ^{12}CO}}{3 k_{\rm B} T_{\rm ex} \Delta v} \exp{\left(\frac{-h \nu_{\rm ^{12}CO} J}{2 k_{\rm B} T_{\rm ex}}\right)} \left\{ 1 - \exp{\left(\frac{-h \nu_{\rm ^{12}CO}}{k_{\rm B} T_{\rm ex}}\right)}\right\},
\end{equation}
where $N_{\rm ^{12}CO}$ is the column density of $^{12}$CO and $\Delta v$ is the line width.
If $h \nu \ll k T_{\rm ex}$, the optical depth can be described as
\begin{equation}
\tau_{\rm ^{12}CO} \propto \frac{N_{\rm ^{12}CO}}{\Delta v\ T_{\rm k}^2}.
\label{eq:tau}
\end{equation}
Therefore, $T_{\rm ^{12}CO}/T_{\rm ^{13}CO}$ ratio is a function of $N_{\rm ^{12}CO}$, $\Delta v$ and $T_{\rm k}$ as well as $R_{12/13}(D)$.

We can estimate the $\tau_{\rm ^{12}CO}$ from the $T_{\rm ^{12}CO}/T_{\rm ^{13}CO}$ ratios of each region, if we assume that the $R_{12/13}$ dependence on the galactocentric distance of NGC~3627 is the same as that of the Galaxy.
In table \ref{tab-7}, the $T_{\rm ^{12}CO}/T_{\rm ^{13}CO}$ ratios estimated by linear fitting in figures \ref{fig:SpectraComparison1213} and \ref{fig:SpectraComparison1213Interarm}, radial extents, expected $R_{12/13}$ and $\tau_{\rm ^{12}CO}$ of each region are presented.
The $T_{\rm ^{12}CO}/T_{\rm ^{13}CO}$ ratios of each region are overplotted on the $\tau_{\rm ^{12}CO}-T_{\rm ^{12}CO}/T_{\rm ^{13}CO}$ plot in figure \ref{fig:tau12CO}.

Figure \ref{fig:tau12CO} and table \ref{tab-7} indicate the variety of $\tau_{\rm ^{12}CO}$ among different regions in NGC~3627.
The bar-end, offset and arm regions seem to have high values of $\tau_{\rm ^{12}CO}$ larger than $\sim3$.
On the other hand, $\tau_{\rm ^{12}CO}$ in the center region, bar region and of the larger FWHM component in the interarm region are expected to be $\sim0.2-0.9$, $\sim0.2-0.7$ and $\sim0.3-1.5$, respectively.
If we adopt the local interstellar medium value of $R_{12/13}\sim68$ \citep{Milam+2005}, the optical depths in these three regions are expected to be $\sim4.0$, $\sim2.9$ and $\sim2.4$.
Therefore, the higher $T_{^{12}\rm{CO}}/T_{^{13}\rm{CO}}$ ratios in the center, bar, and interarm regions are likely due to the lower optical depth of $^{12}$CO than the other regions as far as we assume the Galactic $R_{12/13}(D)$.

\begin{table*}
 \caption{The $T_{\rm ^{12}CO}/T_{\rm ^{12}CO}$ ratios, radial extents, expected $R_{12/13}$ and $\tau_{\rm ^{12}CO}$ of each region.}
 \label{tab-7}
 \begin{minipage}{\textwidth}
 \begin{center}
 \begin{tabular}{@{}lcccccc}
  \hline
 & Center & Bar & Bar-end & Offset & Arm & Interarm$^\ast$\\
  \hline
  \hline
$T_{\rm ^{12}CO}/T_{\rm ^{13}CO}$$^{\ast\ast}$& $17.1$ & $22.7$ & $12.2$ & $11.2$ & $12.9$ & $26.4$\\
$D$ (kpc) & $0-1$ & $1-2$ & $2-5$ & $3-6$ & $3-6$ & $2-5$\\
$R_{12/13}$ & $19-25$ & $25-31$ & $31-50$ & $37-56$ & $37-56$ & $31-50$ \\
$\tau_{\rm ^{12}CO}$ & $0.2-0.9$ & $0.2-0.7$ & $2.4-4.2$ & $3.3-5.2$ & $2.8-4.5$ & $0.3-1.5$\\
  \hline
  \end{tabular}
\end{center}
\footnotetext[$\ast$]{``outskirt'' range.}
\footnotetext[$\ast\ast$]{The slopes obtained in figure \ref{fig:SpectraComparison1213} and \ref{fig:SpectraComparison1213Interarm}.}
\end{minipage}
\end{table*}

\subsubsection{The cause of the low $\tau_{\rm ^{12}CO}$ in the bar and center regions}

According to equation (\ref{eq:tau}), the $T_{\rm ^{12}CO}/T_{\rm ^{13}CO}$ ratio is a function of $R_{12/13}(D)$, $N_{\rm ^{12}CO}$, $\Delta v$ and $T_{\rm k}$.
The physical states of molecular gas in the bar and center regions are discussed in W11\nocite{Watanabe+2011}.
They concluded that the higher value of $I_{\rm ^{12}CO}/I_{\rm ^{13}CO}$ in the center and bar regions are due to differences in the environments:
the high $T_{\rm k}$, which is due to the starburst or nuclear activity in the center region, and the low $N_{\rm CO}/\Delta v$, which is due to a streaming motion in the bar region.
The moderate value of $\tau_{\rm ^{12}CO}$ in the central region has also been reported in the study of our Galaxy \citep{Oka+1998}.

\subsubsection{The possible origin of the low $\tau_{\rm ^{12}CO}$ gas component in the interarm region}

The previous studies of GMCs in the Milky Way and in the extragalactic objects have proposed an evolution scenario of GMCs in a galactic disk.
\citet{Sawada+2012a} investigated the structure and physical conditions of molecular gas in the Milky Way including the arm and interarm regions using data taken with the 45-m telescope at NRO (spatial resolution of $\sim0.5$ pc).
They concluded that when faint and diffuse molecular gas in the interarm region enters the spiral arm, this gas develops bright and compact structures at the arm and once the gas leaves the arms, it returns to a diffuse state \citep{Sawada+2012b}.
\citet{Koda+2009} utilized CARMA data of M~51 (spatial resolution of $\sim160$ pc) and showed that molecular clouds with mass of $\sim 10^{7-8}$ $M_\odot$ are found only in the arm region while one with $\sim 10^{5-6}$ $M_\odot$ can be found in the interarm as well as arm regions.
They claimed that massive clouds that are accumulated in the arm region are not fully dissociated into atomic gas, but dissolved into small clouds as they pass through the arm due to the shear motion in the interarm region.
Recently, \citet{Colombo+2014} confirmed the results of \citet{Koda+2009} quantitatively with a large number of GMC samples of M~51 (spatial resolution of $\sim50$ pc) by comparing the mass function of GMC in different regions.
However, these extragalactic studies mainly treat individual and discrete clouds and might miss the diffuse and extended objects.

Our result suggests the existence of diffuse non-optically thick $^{12}$CO component ($\tau_{\rm ^{12}CO}\sim0.3-1.5$) in the interarm region.
This is consistent with the conclusion given in Sawada et al. (2012a; 2012b)\nocite{Sawada+2012a,Sawada+2012b} although they did not analyze the optical depth of the ``diffuse'' component.
\citet{Polk+1988} also reported that a significant contribution to the large-scale CO emission from the Galaxy is made by diffuse gas, which is indicated from the extremely high $I_{\rm ^{12}CO}/I_{\rm ^{13}CO}$ ratio of $\sim20-50$ \citep{KnappBowers1988}.
The diffuse component observed in our study might be a result of the dissolution of massive molecular clouds in the interarm region, as suggested by \citet{Koda+2009} and \citet{Colombo+2014}.
Relatively low $^{12}$CO optical depth of the interarm region may be a result of low $N_{\rm ^{12}CO}$$/\Delta v$ due to a shear motion.
Therefore, it is possible that GMCs formed at the arm are dissolved into smaller GMCs and diffuse molecular component in the interarm regions.
This picture is consistent with the prediction from a recent numerical simulation of gas component under spiral potential in a disk galaxy \citep{DobbsPringle2013}.
Some of the GMCs at the arm may be dissociated into atomic gas in the interarm region by star-formation feedback (e.g., \cite{Dobbs+2006}).
However, a detailed understanding {\rm of} the mechanism of dissolution of GMC or dissociation of molecular gas requires molecular and atomic gas observations in high resolution and sensitivity with Atacama Large Millimeter/submillimeter Array (ALMA) and Square Kilometer Array (SKA).

There may be a contribution of the bar structure on the low $\tau_{\rm ^{12}CO}$ value in the interarm region, since the interarm region that we define in this study includes the neighboring areas on the sides of the bar region.
The bar consists of a lot of characteristic orbits of stars and one of the most prominent orbits is called the x1 orbit, which shapes the elongated structure.
Since the gas component has viscosity contrary to stars, its orbit around the bar structure deviates from the sequence of x1 orbits of the stellar components.
\citet{Wada1994} provided an analytical model for the orbits of gas component in the bar-structure, damped orbit model.
In his model, some gas orbits in the bar, especially at the dust lane, gradually deviate from the bar region (for example, see figure 10 of \cite{Sakamoto+1999}).
Hence, some gas components orbiting the bar structure may be categorized in this study as interarm gas components.
Furthermore, the orbits which are crowded at the bar-end region become sparse on the sides of the bar region.
The gap between orbits widens, perhaps reducing ${N_{\rm ^{12}CO}}/\Delta v$ and consequently decreasing $\tau_{\rm ^{12}CO}$ on the sides of the bar.
Unfortunately, no study so far has investigated this effect in detail.
The investigation of this effect is an issue for a future paper.

\subsection{Non-universal CO-to-H$_2$ conversion factor in a galaxy?}
\label{subsec:conversionfactor}

Most previous studies of nearby galaxies have adopted a universal conversion factor for the entire galaxy and investigated the distribution of molecular gas and star-formation efficiency (SFE), which is determined as a ratio of star-formation rate and molecular gas mass (e.g., \cite{Helfer+2003,Kuno+2007,Leroy+2009}).
Some studies claimed that the SFE in the bar region is lower than the other regions and suggested that intense phenomena such as a streaming motion inhibits star formation in the bar (e.g., \cite{ReynaudDowns1998}).
W11\nocite{Watanabe+2011} compared SFE obtained from the $^{12}$CO data and from $^{13}$CO data of NGC~3627 and concluded that SFE in the bar region is comparable to that of the arm region if $^{13}$CO data were used to estimate the molecular gas mass.
A lower conversion factor in the bar region than the other disk regions is also suggested in the $^{12}$CO data of Maffei II combined with an LVG (large velocity gradient) analysis \citep{Sorai+2012}.

In this study, we detect $^{13}$CO emission from the interarm region of NGC~3627 for the first time and find $^{12}$CO component with a broad line width and high $T_{\rm ^{12}CO}/T_{\rm ^{13}CO}$ ratio indicating non-optically thickness.
It is difficult to estimate the molecular gas mass in the interarm region only with $^{13}$CO since a part of the molecular gas in the interarm region is expected to be too diffuse to emit $^{13}$CO emission.
To infer the impact of the non-optically thick $^{12}$CO component on the observed $^{12}$CO integrated intensity and the CO-to-H$_2$ conversion factor, we use the stacked $^{12}$CO spectrum to estimate the fraction of the emission from the diffuse component and the molecular gas mass in the interarm region.

\begin{figure*}
\includegraphics[width=140mm]{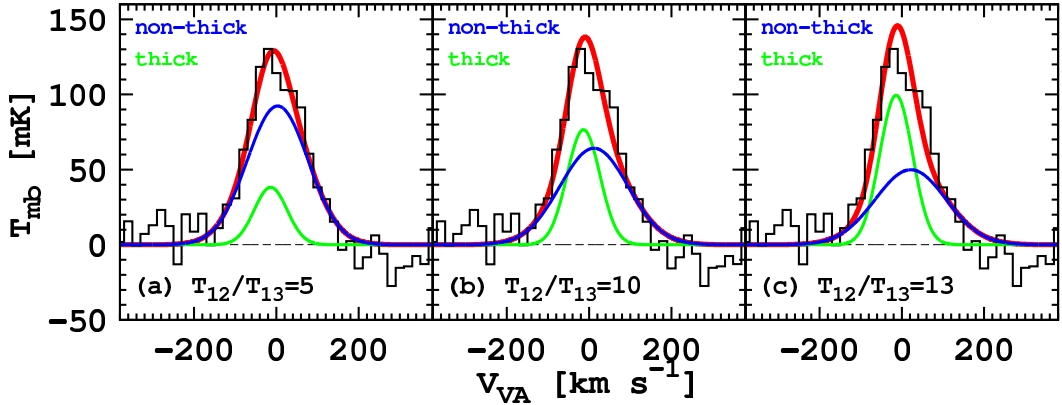}
 \vspace{0cm}
 \caption{Fitting results with two Gaussians of $^{12}$CO spectrum in the interarm region. Blue and green solid lines show the fitting results for non-optically thick and optically thick components, respectively. Red solid line shows the sum of them.}
  \label{fig:DualGaussians}
\end{figure*}

We decomposed the $^{12}$CO spectra with two Gaussians for $^{12}$CO optically thick and non-optically thick components.
For the optically-thick component, we adopted the center velocity of $-13.7$ km s$^{-1}$ and the FWHM of $94.4$ km s$^{-1}$ from the Gaussian fitting results of the $^{13}$CO spectra and assumed three cases of $T_{\rm peak}$ of $T_{\rm ^{12}CO}/T_{\rm ^{13}CO}$ = 5, 10 and 13.
The $T_{\rm ^{12}CO}/T_{\rm ^{13}CO}$ ratios are determined by reference to the typical $I_{\rm ^{12}CO}/I_{\rm ^{13}CO}$ ratio of the Galactic GMCs of $5-7$ \citep{Solomon+1979,Polk+1988} and the values which we found in the bar-end, offset and arm regions of NGC~3627.
As a result, the fractions of the non-optically thick $^{12}$CO emission with respect to the total $^{12}$CO flux, $f_{\rm thin}$ are $82\pm11$ \%, $64\pm12$ \%, and $52\pm13$ \% if we assume $T_{\rm ^{12}CO}/T_{\rm ^{13}CO}$ = 5, 10 and 13, respectively.
The fitting results for these three cases are shown in figure \ref{fig:DualGaussians} and summarized in table \ref{tab-8}.

The surface density of the molecular gas $\Sigma_{\rm mol}$ in the interarm region is described with integrated intensities of the optically thick and thin components, $I_{\rm ^{12}CO, thick}$ and $I_{\rm ^{12}CO, thin}$ using equation (\ref{eq:mol12thick}) and (\ref{eq:mol12thin}) as\footnote{
It is reported that $N_{\rm ^{12}CO}/N_{\rm H_2}$ in the diffuse gas is smaller than the one in the GMCs from the CO-absorption observations \citep{Sonnentrucker+2007,Liszt2007,Burgh+2007,Shetty+2008}.
However, the $N_{\rm ^{12}CO}/N_{\rm H_2}$ values reported in those studies have a large dispersion ($\sim 10^{-7}-10^{-5}$).
Therefore we adopt the GMC value of $10^{-4}$ \citep{YoungScoville1991} throughout this paper.
}
\begin{equation}
\Sigma_{\rm mol,interarm}\approx 1.34\ I_{\rm ^{12}CO, thick} + 0.133\ I_{\rm ^{12}CO, thin}\ \ \ M_\odot\ \rm{pc^{-2}}.
\label{eq-11}
\end{equation}
We obtain the surface densities of $7.4\pm2.3$ $M_\odot$ pc$^{-2}$ ($T_{\rm ^{12}CO}/T_{\rm ^{13}CO}$ = 5), $12.0\pm2.6$ $M_\odot$ pc$^{-2}$ (10) and $14.8\pm3.0$ $M_\odot$ pc$^{-2}$ (13), which are significantly lower than the one calculated by assuming all $^{12}$CO emission is optically thick by factors of $3.8\pm1.2$, $2.3\pm0.6$ and $1.9\pm0.5$, respectively.
Here we assume $N_{\rm ^{12}CO}/N_{\rm H_2}=10^{-4}$ and $T_{\rm k}=20$ K.
It should be noted that this is an extreme case where the non-optically thick component is optically thin ($\tau_{\rm ^{12}CO}\ll 1$) so these factors set only the upper limits.

A radial gradient of $X_{\rm CO}$ in nearby galaxies has been investigated observationally, but the consensus has not been obtained \citep{Sandstrom+2013,Blanc+2013}.
\citet{Blanc+2013} investigated the dependence of $X_{\rm CO}$ of NGC~628 on the metallicity, gas surface density, and UV radiation field, which all affect the balance between the shielding and dissociation of CO molecule in the photodissociation regions on the edge of molecular clouds.

The conversion factor is expected to vary according to not only the radiative transfer of UV between the star-forming region and the subjected clouds in the galaxy but also the radiative transfer of $^{12}$CO between the clouds and us.
The previous studies have mainly focused on the factors that affect the former radiative transfer and investigated the dependence of $X_{\rm CO}$ on the metallicity, gas surface density, and UV radiation field.
Here we show that the optical depth of $^{12}$CO that influences the latter radiative transfer may be different in the different regions in NGC~3627, resulting in the different conversion factor.
We need careful consideration and treatment when we estimate the molecular gas mass not only in the region with quite different metallicity, ISM density and UV radiation field but also in the region where $^{12}$CO is not expected to be optically thick.
Accurate estimation of the molecular gas mass is also important to evaluate other physical parameters of galaxies such as SFE.

\begin{table*}
\caption{Integrated intensity estimated with a dual-Gaussian fitting$^\ast$ to the stacked $^{12}$CO spectra of the interarm region.}
 \begin{minipage}{\textwidth}
 \begin{center}
 \begin{tabular}{lcccc}
  \hline
$T_{\rm ^{12}CO}/T_{\rm ^{13}CO}$ & $T_{\rm peak, ^{12}CO, thick}$ & $I_{\rm ^{12}CO, thick}$ & $I_{\rm ^{12}CO, thin}$ & $f_{\rm thin}$\\
&  (mK) &  (K km s$^{-1}$) &  (K km s$^{-1}$) & (\%)\\
\hline
\hline
5 & $38.2\pm4.5$ & $3.8\pm0.7$ & $17.2\pm1.7$ & $82\pm11$\\
10 & $76.5\pm9.0$ & $7.7\pm1.4$ & $13.3\pm2.0$ & $64\pm12$\\
13 & $99.4\pm11.7$ & $10.0\pm1.8$ & $10.8\pm2.2$ & $52\pm13$\\
\hline
\end{tabular}
\end{center}
\footnotetext[$\ast$]{
Fitting with two Gaussians: one is for the optically thick component and the other for the non-optically thick component.
The free parameters for fitting are $T_{\rm peak}$, line center velocity and FWHM for the non-optically thick component.
The center velocity and FWHM for the optically thick component are fixed to $-13.7$ km s$^{-1}$ and 94.1 km s$^{-1}$, respectively, which are estimated by fitting the $^{13}$CO spectra with a Gaussian.
}
\end{minipage}
\label{tab-8}
\end{table*}%

\section{Summary}\label{Summary}
We obtained the averaged spectra of $^{12}$CO and $^{13}$CO in the center, bar, bar-end, offset, arm and interarm regions of NGC~3627 with the stacking analysis after the velocity-axis alignment (VA) procedure according to the velocity field estimated from the $^{12}$CO mapping data.
The $^{13}$CO spectrum in the interarm region of NGC~3627 where the emission does not have enough S/N in the original data was successfully detected.
Main results of this paper are as follows:

\begin{enumerate}
\item A weak $^{13}$CO emission in the interarm region of NGC~3627 is successfully detected for the first time with the stacking analysis after the VA procedure (figure \ref{fig:StackedSpectra} of section \ref{Results}).

\item The validity of the stacking method with VA is confirmed by comparing the integrated intensity between stacked spectra with and without the VA procedure.
Moreover, the S/N of stacked spectra with VA is improved by a factor of up to 3.2 compared to those without VA (table \ref{tab-2} of section \ref{Results}).

\item The integrated intensity ratios $I_{^{12}\rm{CO}}/I_{^{13}\rm{CO}}$ in the bar and interarm regions are almost two times higher than those in the other regions.
$I_{^{12}\rm{CO}}/I_{^{13}\rm{CO}}$ in the center region is the intermediate value between them.
High values of $I_{^{12}\rm{CO}}/I_{^{13}\rm{CO}}$ in the bar and center regions are attributed to the higher intensity ratios ($T_{\rm ^{12}CO}/T_{\rm ^{13}CO}$) and one in the interarm regions is attributed to the higher FWHM$_{^{12}\rm{CO}}/$FWHM$_{^{13}\rm{CO}}$ ratio than the other regions.
The difference in the line width between $^{12}$CO and $^{13}$CO suggests two gas components, one with a narrow ($\sim$ FWHM$_{\rm ^{13}CO}$) and the other with a broad line width ($\sim$ FWHM$_{\rm ^{12}CO}$) in the interarm region (tables \ref{tab-4} and \ref{tab-5} of section \ref{subsec-Discussion2}).

\item $T_{\rm ^{12}CO}/T_{\rm ^{13}CO}$ in the center and bar regions and of the broad line width components in the interarm region are $17.1$, $22.7$ and $26.4$ indicating that the $^{12}$CO lines are not completely optically thick in those regions if we assume the same $^{12}$C/$^{13}$C radial gradient as that of our Galaxy
(figures \ref{fig:SpectraComparison1213}, \ref{fig:SpectraComparison1213Interarm} and \ref{fig:tau12CO} of sections \ref{subsec:slope} and \ref{subsec:physicalstate}).

\item More than half of the $^{12}$CO emission from the interarm region is likely to be radiated from the diffuse gas component, if the $^{12}$CO spectra is decomposed with two Gaussians, one with FWHM$_{^{13}\rm{CO}}$ and the other with $\sim$ FWHM$_{^{12}\rm{CO}}$ (figure \ref{fig:DualGaussians} of section \ref{subsec:conversionfactor}).

\item The existence of non-optically thick component of $^{12}$CO in the center, bar, and interarm regions indicates a lower CO-to-H$_2$ conversion factor compared to the other regions.
It is necessary to take into account the non-universal conversion factor in a galaxy in case of comparing the molecular gas distribution and SFE in the different regions.
Otherwise, the molecular gas mass and SFE may be respectively overestimated and underestimated by factors of a few in case of the interarm region of NGC~3627 (section \ref{subsec:conversionfactor}).

\end{enumerate}




\bigskip

We would like to thank an anonymous referee for very productive comments.
KMM thanks Shuuro Takano, Tetsuhiro Minamidani, Tomoki Morokuma, Junichi Baba, Daisuke Iono, Jin Koda and all members of NRO for their support and fruitful discussions.

This research has made use of the NASA/ IPAC Infrared Science Archive, which is operated
by the Jet Propulsion Laboratory, California Institute of Technology, under contract with the
National Aeronautics and Space Administration.






\begin{thebibliography}{}
\bibitem[Bally \& Langer(1982)]{BallyLanger1982}
  Bally J. \& Langer W. D., 1982, \apj, 255, 143
\bibitem[Blanc et al.(2013)]{Blanc+2013}
  Blanc G. A. et al., 2013, \apj, 764, 117
\bibitem[Bolatto et al.(2013)]{Bolatto+2013}
  Bolatto A. D. et al., 2013, \araa, 51, 207
\bibitem[Burgh et al.(2007)]{Burgh+2007}
  Burgh E. B. et al., 2007, \apj, 658, 446
\bibitem[Colombo et al.(2014)]{Colombo+2014}
  Colombo D. et al., 2014, \apj, 784, 3
\bibitem[de Vaucouleurs et al.(1991)]{deVaucouleurs+1991}
  de Vaucouleurs G. et al., 1991, Third Reference Catalogue of Bright Galaxies. Volume I: Explanations and references. Volume II: Data for galaxies between 0$^h$ and 12$^h$. Volume III: Data for galaxies between 12$^h$ and 24$^h$.
\bibitem[Dobbs et al.(2006)]{Dobbs+2006}
  Dobbs, C. L. et al., 2006, \mnras, 371, 1663
\bibitem[Dobbs \& Pringle(2013)]{DobbsPringle2013}
  Dobbs, C. L. \& Pringle J. E., 2013, \mnras, 432, 653
\bibitem[Egusa et al.(2011)]{Egusa+2011}
  Egusa, F. et al., 2011, \apj, 726, 85
\bibitem[Fujimoto(1968)]{Fujimoto1968}
  Fujimoto, M, 1968, in IAU Symp. 29, Non-stable Phenomena in Galaxies, (Yerevan: The Publishing House of the Academy of Sciences of Armenian SSR), 453
\bibitem[Frerking et al.(1982)]{Frerking+1982}
  Frerking M. A. et al., 1982, \apj, 262, 590
\bibitem[Galametz et al.(2012)]{Galametz+2012}
  Galametz M. et al., 2012, \mnras, 425, 763
\bibitem[Haynes et al.(1979)]{Haynes+1979}
  Haynes M. P. et al., 1979, \apj, 229, 83
\bibitem[Helfer et al.(2003)]{Helfer+2003}
  Helfer T. T. et al., 2003, \apjs, 145, 259
\bibitem[Hirota et al.(2010)]{Hirota+2010}
  Hirota A. et al., 2010, \pasj, 62, 1261
\bibitem[H\"{u}ttemeister et al.(2000)]{Huttemeister+2000}
  H\"{u}ttemeister S. et al., 2000, \aap, 363, 93
\bibitem[Kennicutt et al.(2003)]{Kennicutt+2003}
  Kennicutt R. C. Jr. et al., 2003, \pasp, 115, 928
\bibitem[Knapp \& Bowers(1988)]{KnappBowers1988}
  Knapp G. R. \& Bowers P. F., 1988, \apj, 331, 974
\bibitem[Koda et al.(2009)]{Koda+2009}
  Koda J. et al., 2009, \apjl, 700, L132
\bibitem[Kopp et al.(1996)]{Kopp+1996}
  Kopp M. et al., 1996, \aap, 305, 558
\bibitem[Krips et al.(2008)]{Krips+2008}
  Krips M. et al., 2008, \apj, 677, 262
\bibitem[Kuno et al.(2007)]{Kuno+2007}
  Kuno N. et al., 2007, \pasj, 59, 117 (K07)
\bibitem[Langer \& Penzias(1990)]{LangerPenzias1990}
  Langer W. D. \& Penzias A. A., 1990, \apj, 357, 477
\bibitem[Leroy et al.(2009)]{Leroy+2009}
  Leroy A. K. et al., 2009, \aj, 137, 4670
\bibitem[Liszt(2007)]{Liszt2007}
  Liszt H. S., 2007, \aap, 476, 291
\bibitem[Meier \& Turner(2004)]{MeierTurner2004}
  Meier D. S. \& TurnerJ. L., 2004, \aj, 127, 2069
\bibitem[Milam et al.(2005)]{Milam+2005}
  Milam S. N. at al., 2005, \apj, 634, 1126
\bibitem[Nakai \& Kuno(1995)]{NakaiKuno1995}
  Nakai N. \& Kuno N., 1995, \pasj, 47, 761
\bibitem[Oka et al.(1998)]{Oka+1998}
  Oka T. et al., 1998, \apjs, 118, 455
\bibitem[Polk et al.(1988)]{Polk+1988}
  Polk K. S. et al., 1988, \apj, 332, 432  
\bibitem[Reynaud \& Downs(1998)]{ReynaudDowns1998}
  Reynaud D. \& Down D., 1998, \aap, 337, 671
\bibitem[Roberts(1969)]{Roberts1969}
  Roberts, W. W. 1969, \apj, 158, 123
\bibitem[Paglione et al.(2001)]{Paglione+2001}
  Paglione T. A. D. et al., 2001, \apjs, 135, 183
\bibitem[Saha et al.(1999)]{Saha+1999}
  Saha A. et al., 1999, \apj, 522, 802
\bibitem[Sakamoto et al.(1999)]{Sakamoto+1999}
  Sakamoto K. et al., 1999, \apjs, 124, 403
\bibitem[Sawada et al.(2012a)]{Sawada+2012a}
  Sawada T. et al., 2012a, \apj, 752, 118
\bibitem[Sawada et al.(2012b)]{Sawada+2012b}
  Sawada T. et al., 2012b, \apj, 759, 26
\bibitem[Sanders et al.(1985)]{Sanders+1985}
  Sanders D. B. et al., 1985, \apj, 289, 373
\bibitem[Sandstrom et al.(2013)]{Sandstrom+2013}
  Sandstrom K. M. et al., 2013, \apj, 777, 5
\bibitem[Schruba et al.(2011)]{Schruba+2011}
  Schruba A. et al., 2011, \aj, 142, 37
\bibitem[Schruba et al.(2012)]{Schruba+2012}
  Schruba A. et al., 2012, \aj, 143, 138
\bibitem[Scoville et al.(1987)]{Scoville+1987}
  Scoville N. Z. et al., 1987, \apjs, 63, 821
\bibitem[Shetty et al.(2008)]{Shetty+2008}
  Shetty Y. et al., 2008, \apj, 687, 1075
\bibitem[Smith \& Adams(1980)]{SmithAdams1980}
  Smith D. \& Adams N. G., 1980, \apj, 242, 424
\bibitem[Sonnentrucker et al.(2007)]{Sonnentrucker+2007}
  Sonnentrucker P. et al., 2007, \apjs, 168, 58
\bibitem[Sorai et al.(2000)]{Sorai+2000}
  Sorai K. et al., 2000, \procspie, 4015, 86
\bibitem[Sorai et al.(2012)]{Sorai+2012}
  Sorai K. et al., 2012, \pasj, 64, 51
\bibitem[Solomon et al.(1979)]{Solomon+1979}
  Solomon P. M. et al., 1979, \apj, 232, 89
\bibitem[Solomon et al.(1987)]{Solomon+1987}
  Solomon P. M. et al., 1987, \apj, 319, 730
\bibitem[Solomon et al.(1997)]{Solomon+1997}
  Solomon P. M. et al., 1997, \apj, 478, 144
\bibitem[van Dishoeck et al.(1988)]{vanDishoeck+1988}
  van Dishoeck E. F. et al., 1988, \apj, 334, 771
\bibitem[Tosaki et al.(2002)]{Tosaki+2002}
  Tosaki T. et al., 2002, \pasj, 54, 209
\bibitem[Wada(1994)]{Wada1994}
  Wada K., 1994, \pasj, 165, 172
\bibitem[Watanabe et al.(2011)]{Watanabe+2011}
  Watanabe Y. et al., 2011, \mnras, 411, 1409 (W11)
\bibitem[Watson et al.(1976)]{Watson+1976}
  Watson W. D. et al., 1976, \apj, 205, 165
\bibitem[Young \& Scoville(1991)]{YoungScoville1991}
  Young J. S. \& Scoville N. Z. et al., 1991, \araa, 29, 581
\bibitem[Zhang et al.(1993)]{Zhang+1993}
  Zhang X. et al., 1993, \apj, 418, 100
\end{thebibliography}
\end{document}